\documentclass[11pt,showpacs,preprintnumbers,superscriptaddress,amsmath,amssymb,nofootinbib]{revtex4}

\usepackage{graphicx}
\usepackage{dcolumn}
\usepackage{bm}
\usepackage{amssymb}
\usepackage{amsmath}
\usepackage{epsfig}    
\usepackage{color}
\usepackage{slashed}

\begin{document} 
\title{Implications of Higgs boson search data on the two-Higgs doublet models with a softly broken $Z_2$ symmetry}

\author{Cheng-Wei Chiang}
\email{chengwei@ncu.edu.tw}
\affiliation{Department of Physics and Center for Mathematics and Theoretical Physics, 
National Central University, Chungli, Taiwan 32001, ROC}
\affiliation{Institute of Physics, Academia Sinica, Taipei, Taiwan 11529, ROC}
\affiliation{Physics Division, National Center for Theoretical Sciences, Hsinchu, Taiwan 30013, ROC}
\author{Kei Yagyu}
\email{keiyagyu@ncu.edu.tw}
\affiliation{Department of Physics and Center for Mathematics and Theoretical Physics,
National Central University, Chungli, Taiwan 32001, ROC}

\begin{abstract}
Based on current data of the Higgs boson search at the Large Hadron Collider, we constrain the parameter space of the two-Higgs doublet models where a softly broken $Z_2$ symmetry is employed to avoid flavor-changing neutral currents at tree level.  There are four types of Yukawa interactions under the $Z_2$ charge assignments of the standard model fermions.  We find that the model with Type-II Yukawa interactions can better explain the experimental data among all.
In this scenario, the couplings of the light CP-even Higgs boson $h$ with weak gauge bosons are almost standard model-like or only slightly different in a small range of $\tan\beta$.  In particular, we scrutinize a well-constrained region previously ignored by other analyses and study the phenomenology of the extra Higgs bosons at the Large Hadron Collider. 
\end{abstract}
\maketitle
\newpage

\section{Introduction}

In the standard model (SM) of particle physics, the Higgs field plays a central role in breaking the electroweak symmetry and giving masses to other elementary particles.  A direct consequence is the existence of a spin-0 Higgs boson that interacts with the other particles with strengths proportional to their respective masses.  A new particle, denoted by $h$, with mass about $125.5$ GeV has been recently discovered at the Large Hadron Collider (LHC) by both the ATLAS and CMS Collaborations~\cite{PLB_ATLAS,PLB_CMS}, and is found to be compatible with the SM Higgs boson in the production rates of various channels.  Moreover, the diphoton decay of $h$ suggests that its spin is an even integer~\cite{LandauYang}, and a preliminary angular analysis of the $h \to ZZ^* \to 4\ell$ decay data prefers that the new boson has zero spin and positive parity~\cite{Angular}.

Even though the new particle $h$ is generally consistent with the SM Higgs boson, one obvious question is whether it is the only spin-0 particle in the SM or actually belongs to a larger scalar family.  There are many well-motivated models with an extended Higgs sector, among which the two-Higgs doublet model (THDM) is one of the most popular and extensively analyzed classes.  Such a structure of two Higgs doublet fields is also required for the Minimal Supersymmetric Standard Model (MSSM). 
By introducing a cousin doublet of the SM Higgs doublet, the model preserves the custodial symmetry in the kinetic term 
to keep the electroweak $\rho$ parameter at unity at tree level.

Recently, there are many analyses about the THDM's~\cite{2gamma_THDM}, with particular emphasis on the study of enhancement in the $h\to \gamma\gamma$ mode as the data show.  A survey of generic models with an extended Higgs sector, including the THDM's, has been done by the authors~\cite{Chiang:2012qz} to show the correlation between the $\gamma\gamma$ and $Z\gamma$ modes.  Obviously, a global analysis to the available data is required to disentangle among the possible new physics candidates~\cite{Low}.
Various scenarios in the THDM have been analyzed by using the Higgs boson search data at the LHC 
before the Moriond conference~\cite{Cheon_Kang, THDMX, Gori, Lee2, Kingman,Drozd, Ferreira_Haber_Santos,Chen_Dawson}, and those afterwards~\cite{After_Moriond}.

In the THDM's, there may be tree-level flavor-changing neutral current (FCNC) interactions due to the mediations of neutral scalar bosons as both doublet fields can generally couple to the up-type and down-type quarks and charged leptons.  There are several ways to avoid such dangerous FCNC processes, {\it e.g.}, imposing a softly broken $Z_2$ symmetry~\cite{GW}\footnote{The model with an exact $Z_2$ symmetry is known as the inert doublet model~\cite{id}.} or 
assuming alignments in the Yukawa matrices~\cite{pich}.  In the former case, there are four independent types of Yukawa interactions, depending on the charge assignments 
of the SM fermions under the $Z_2$ symmetry~\cite{z2}.  They are dubbed the Type-I, Type-II, Type-X and Type-Y THDM's~\cite{typeX,typeX2}. 
After electroweak symmetry breaking, 
all these scenarios contain the same physical Higgs bosons under the assumption of CP conservation in the Higgs sector: two charged Higgs bosons, two CP-even Higgs bosons, and one CP-odd Higgs boson.

The Type-II THDM can best fit the latest Higgs search data from the LHC, with the preferred values $\tan\beta \sim 4$ where $\tan\beta$ is defined as the ratio of the vacuum expectation values (VEV's) of the two doublet fields and SM-like couplings among the lighter CP-even Higgs boson and weak gauge bosons $g_{hVV}$~\cite{Lee2,Kingman}. 
However, we find that there is another parameter region that cannot be ruled out by the data at even $68\%$ confidence level (CL) and is still consistent with the constraints of vacuum stability and perturbative unitarity.  
This region corresponds to the case where the deviation in the $g_{hVV}$ coupling is more than $-1\%$ from its SM value, and the values of the tau and bottom Yukawa couplings are slightly smaller in magnitude and have the opposite sign to their SM values. 
Although the Higgs boson signal are still SM-like, such a case has effects on the productions and decays of the other Higgs bosons at colliders.  In this paper, we study the Higgs phenomenology at the LHC in this part of the parameter space.

This paper is organized as follows.  In Section~\ref{sec:THDM}, we review the THDM with a $Z_2$ symmetry and classify four types of Higgs interactions with SM fermions.  A soft $Z_2$ symmetry breaking term is also introduced.  In Section~\ref{sec:fit}, we perform $\chi^2$ fits of the four scenarios to the current data, singling out the Type-II interactions as the preferred one.  We then find two separate regions in the model parameter space, noting that one of them has been ignored in previous analyses.  We then concentrate on the region where the $g_{hVV}$ couplings have larger deviations from the SM expectations, and study the collider phenomenology of the heavier Higgs bosons in Section~\ref{sec:pheno}.  After working out the branching fractions of both heavy CP-even Higgs boson, $H$, and CP-odd Higgs boson, $A$, we examine their single and pair productions at the LHC and compare the results with current search limits.  Our findings are summarized in Section~\ref{sec:summary}.

\section{The two-Higgs doublet model \label{sec:THDM}}

The THDM contains two $SU(2)_L$ doublet Higgs fields $\Phi_1$ and $\Phi_2$ with hypercharge $Y=+1/2$. 
In general, both doublet fields can couple to the SM fermions at the same time and induce FCNC's via the mediation of a scalar boson at tree level. 
To avoid such FCNC's, a softly broken discrete $Z_2$ symmetry is introduced to the model,
under which the doublet fields transform as $\Phi_1\to +\Phi_1$ and $\Phi_2\to -\Phi_2$. 
\begin{table}[t]
\begin{center}
\begin{tabular}{|c||c|c|c|c|c|c||c|c|c|}
\hline & $\Phi_1$ & $\Phi_2$ & $u_R^{}$ & $d_R^{}$ & $\ell_R^{}$ &
 $Q_L$, $L_L$ &$\xi_u$ & $\xi_d$ & $\xi_e$\\  \hline
Type-I  & $+$ & $-$ & $-$ & $-$ & $-$ & $+$ & $\cot\beta$ & $\cot\beta$ & $\cot\beta$\\
Type-II & $+$ & $-$ & $-$ & $+$ & $+$ & $+$ & $\cot\beta$ & $-\tan\beta$ & $-\tan\beta$\\
Type-X  & $+$ & $-$ & $-$ & $-$ & $+$ & $+$ & $\cot\beta$ & $\cot\beta$ & $-\tan\beta$\\
Type-Y  & $+$ & $-$ & $-$ & $+$ & $-$ & $+$ & $\cot\beta$ & $-\tan\beta$ & $\cot\beta$\\
\hline
\end{tabular} 
\end{center}
\caption{Charge assignments of the $Z_2$ symmetry and the corresponding $\xi_f$ factors in different scenarios of the THDM.} \label{Tab:type}
\end{table}
%
%
%
The two doublet fields can be parameterized in the so-called Higgs basis as
\begin{align}
\left(\begin{array}{c}
\Phi_1\\
\Phi_2
\end{array}\right)=R(\beta)\left(\begin{array}{c}
\Phi\\
\Psi
\end{array}\right),~~\text{with}~~
R(\theta)=\left(\begin{array}{cc}
\cos\theta & -\sin\theta\\
\sin\theta & \cos\theta
\end{array}\right), 
\label{gb11}
\end{align}
where
\begin{align}
\Phi=\left[
\begin{array}{c}
w^+\\
\frac{1}{\sqrt{2}}(v+h_1'+iz)
\end{array}\right],\quad
\Psi=\left[
\begin{array}{c}
H^+\\
\frac{1}{\sqrt{2}}(h_2'+iA)
\end{array}\right]
\text{ and }
\tan\beta =\langle\Phi_2^0\rangle/\langle \Phi_1^0 \rangle , \label{gb22}
\end{align}
with the VEV $v = \sqrt{\langle\Phi_1^0\rangle^2 + \langle \Phi_2^0 \rangle^2} = 246$ GeV.
In the above parameterization of the component scalar fields, $w^\pm$ and $z$ are the Nambu-Goldstone bosons becoming the longitudinal components of the $W^\pm$ and $Z$ bosons, respectively. 
The physical charged Higgs boson (the CP-odd Higgs boson) is denoted by $H^\pm$ ($A$) and 
the CP-even Higgs bosons are expressed as $h_1'$ and $h_2'$. 
In general, ($h_1'$, $h_2'$) are not mass eigenstates and can mix with each other.  
The mass eigenstates of the CP-even Higgs bosons can be defined by introducing the mixing angle $\alpha$ as~\footnote{The CP-even scalar states in the basis of ($h_1,h_2$), which are the real parts of the neutral states in $\Phi_1$ and $\Phi_2$, respectively, can be directly related to those in the basis of ($H,h$) by $(h_1,h_2)^T=R(\alpha)(H,h)^T$.}
\begin{align}
\left(\begin{array}{c}
h_1' \\
h_2'
\end{array}\right)= R(\alpha-\beta)
\left(\begin{array}{c}
H\\
h
\end{array}\right). 
\end{align}
We will assume that $h$ is the newly discovered particle and is the lighter mass eigenstate; {\it i.e.}, $h$ is considered as the SM-like Higgs boson, and $H$ is the heavier one.

In the Higgs basis, the Yukawa Lagrangian with the $Z_2$ symmetry are given as
\begin{align}
\mathcal{L}_Y&=
-\frac{\sqrt{2}}{v}\left[\bar{Q}_LM_d\left(\Phi +\xi_d\Psi\right)d_R
+\bar{Q}_LM_u\left(\tilde{\Phi}+\xi_u\tilde{\Psi}\right)u_R
+\bar{L}_LM_e\left(\Phi+\xi_e\Psi\right) e_R
+\text{h.c.}\right], 
\label{Yukawa_xi}
\end{align}
where $\xi_f$ ($f=u,d$ or $e$) can be determined when we specify the $Z_2$ charges of the quarks and leptons, and $M_f$ is the diagonalized fermion mass matrix.  
The charge conjugation of the scalar fields are denoted as $\tilde{\Phi}=i\tau_2\Phi^*$ and $\tilde{\Psi}=i\tau_2\Psi^*$, where $\tau_2$ is the second Pauli matrix.
There are four independent charge assignments of the $Z_2$ parity as summarized in TABLE~\ref{Tab:type}. 
The Yukawa interactions are expressed in terms of the CP-even Higgs mass eigenstates as
\begin{align}
{\mathcal L}_Y
&=
-\sum_{f=u,d,e}\frac{m_f}{v}\Big\{ [\sin(\beta-\alpha)+\xi_f\cos(\beta-\alpha)]{\overline
f}fh\notag\\
&+[\cos(\beta-\alpha)-\xi_f\sin(\beta-\alpha)]{\overline f}fH+i\text{Sign}(f) \xi_f{\overline f}\gamma_5fA\Big\}\notag\\
&-\left[\frac{\sqrt2V_{ud}}{v}\overline{u}
\left( m_d\xi_d P_R - m_u\xi_u P_L \right)d\,H^+
+\frac{\sqrt2m_e\xi_e}{v}\overline{\nu^{}}P_Re^{}H^+
+\text{h.c.}\right], 
\end{align}
where $P_{L,R}$ are the projection operators for left- and right-handed fermions, respectively, and Sign$(f)=+1$ ($-1$) for $f=d$ and $e$ ($f=u$).

We consider the Higgs potential
\begin{align}
V&=m_1^2|\Phi_1|^2+m_2^2|\Phi_2|^2-m_3^2(\Phi_1^\dagger \Phi_2 +\text{h.c.})\notag\\
& +\frac{1}{2}\lambda_1|\Phi_1|^4+\frac{1}{2}\lambda_2|\Phi_2|^4+\lambda_3|\Phi_1|^2|\Phi_2|^2+\lambda_4|\Phi_1^\dagger\Phi_2|^2
+\frac{1}{2}\lambda_5[(\Phi_1^\dagger\Phi_2)^2+\text{h.c.}], \label{pot_thdm2}
\end{align}
where $m_3^2$ and $\lambda_5$ are generally complex parameters.  However, we assume CP-conserving THDM and, therefore, $m_3^2$ and $\lambda_5$ are taken to be real. 
Two of the parameters in the above potential $m_1^2$ and $m_2^2$ can be related to other parameters using the tadpole conditions:
\begin{align}
m_1^2&=m_3^2\tan\beta-\frac{v^2}{2}(\lambda_1\cos^2\beta+\bar{\lambda}\sin^2\beta),\\
m_2^2&=m_3^2\cot\beta-\frac{v^2}{2}(\lambda_2\sin^2\beta+\bar{\lambda}\cos^2\beta),
\end{align}
where $\bar{\lambda}=\lambda_3+\lambda_4+\lambda_5$.  
The masses of $H^\pm$ and $A$ are
\begin{align}
m_{H^+}^2=M^2-\frac{v^2}{2}(\lambda_4+\lambda_5),\quad m_A^2&=M^2-v^2\lambda_5, \label{mass1}
\end{align}
where
\begin{align}
M^2=\frac{m_3^2}{\sin\beta\cos\beta}. \label{bigm}
\end{align}
The masses of the CP-even Higgs bosons are calculated as
\begin{align}
m_H^2&=\cos^2(\alpha-\beta)M_{11}^2+\sin^2(\alpha-\beta)M_{22}^2+\sin2(\alpha-\beta)M_{12}^2,\\
m_h^2&=\sin^2(\alpha-\beta)M_{11}^2+\cos^2(\alpha-\beta)M_{22}^2-\sin2(\alpha-\beta)M_{12}^2, 
\end{align}
where $M_{11}^2$, $M_{22}^2$ and $M_{12}^2$ are the elements of the mass matrix in the basis of ($h_1',h_2'$) expressed by
\begin{subequations}
\begin{align}
M_{11}^2&=v^2(\lambda_1\cos^4\beta+\lambda_2\sin^4\beta)+\frac{v^2}{2}\bar{\lambda}\sin^22\beta,\\
M_{22}^2&=M^2+v^2\sin^2\beta\cos^2\beta(\lambda_1+\lambda_2-2\bar{\lambda}),\\
M_{12}^2&=\frac{v^2}{2}\sin2\beta(-\lambda_1\cos^2\beta+\lambda_2\sin^2\beta)+\frac{v^2}{2}\sin2\beta\cos2\beta\bar{\lambda}, 
\end{align}
\label{mateven}
\end{subequations}
and the mixing angle is given as
\begin{align}
\tan 2(\alpha-\beta)=\frac{2M_{12}^2}{M_{11}^2-M_{22}^2}.
\end{align} 
We note that the decoupling limit can be taken by letting $M^2\to \infty$~\cite{dec_THDM}.  Among $M_{ij}^2$ only $M_{22}^2$ depends on $M^2$, so that $M_{22}^2$ goes to infinity in this limit, corresponding to $\sin(\alpha-\beta)\to -1$. 

It may be useful to write down the explicit formula for the $hH^+H^-$ vertex, 
which is important when we consider the $H^\pm$ loop contribution to the $h\to \gamma\gamma$ decay, as
\begin{align}
\lambda_{hH^+H^-}&=\frac{1}{v\sin2\beta}\left[2\cos(\alpha+\beta)(m_h^2-M^2)+\sin(\beta-\alpha)\sin2\beta(2m_{H^+}^2-m_h^2)\right].
\label{hchch}
\end{align}
We note that when the sign of $\lambda_{hH^+H^-}$ is negative (positive), the $H^\pm$ loop contribution to $h\to \gamma\gamma$ has constructive (destructive) interference with the $W$ boson loop contribution.

The kinetic terms of the Higgs fields in the Higgs basis is given by
\begin{align}
\mathcal{L}_{\text{kin}}&=|D_\mu\Phi|^2+|D_\mu\Psi|^2.  \label{kin1}
\end{align}
where the covariant derivative
\begin{align}
D_\mu = \partial_\mu-\frac{i}{2}g\vec{\tau}\cdot\vec{A}_\mu-\frac{i}{2}g'B_\mu. 
\end{align}
From this equation, it is seen that the first term is identical to the kinetic term of the SM Higgs boson.  Therefore, the weak gauge boson masses are derived in the same way as in the SM.  
The couplings among the CP-even Higgs bosons and the weak gauge bosons, however, can be different from those in the SM, $g_{hVV}^{\text{SM}}$, because of the mixing between the two CP-even states. 
Explicitly,
\begin{align}
g_{hVV} = g_{hVV}^{\text{SM}}\sin(\beta-\alpha),\quad  g_{HVV} = g_{hVV}^{\text{SM}}\cos(\beta-\alpha). 
\end{align}

\section{Data Fitting \label{sec:fit}}

\begin{table}[t]
\begin{center}
\begin{tabular}{|l||c|c||c|c|}\hline 
Mode& Data (ATLAS)&$\int\mathcal{L}$: 7 TeV+8 TeV (ATLAS)& Data (CMS) & $\int\mathcal{L}$: 7 TeV+8 TeV (CMS) \\\hline\hline
$h\to \gamma\gamma$&$1.65^{+0.24+0.25}_{-0.24-0.18}$~\cite{ATLAS_Moriond1}&4.8 fb$^{-1}+20.7$ fb$^{-1}$&$0.77\pm 0.27$ (MVA)~\cite{CMS_Moriond_MVA}  & 5.1 fb$^{-1}+19.6$ fb$^{-1}$  \\\hline
&&&$1.11\pm 0.31$ (Cut Based)~\cite{CMS_Moriond1} & 5.1 fb$^{-1}+19.6$ fb$^{-1}$  \\\hline
$h\to ZZ$&$1.5\pm0.4$~\cite{ATLAS_Moriond2}&4.8 fb$^{-1}+20.7$ fb$^{-1}$& $0.91^{+0.3}_{-0.24}$~\cite{CMS_Moriond2} & 5.1 fb$^{-1}+19.6$ fb$^{-1}$   \\\hline
$h\to WW$&$1.01\pm 0.31$~\cite{ATLAS_Moriond3}&4.8 fb$^{-1}+20.7$ fb$^{-1}$& $0.76\pm 0.21$~\cite{CMS_Moriond3} & 4.9 fb$^{-1}+19.5$ fb$^{-1}$ \\\hline
$h\to b\bar{b}$&$-0.4\pm 1.0$~\cite{ATLAS_Moriond4}&4.7 fb$^{-1}+13$ fb$^{-1}$ & $1.3^{+0.7}_{-0.6}$~\cite{CMS_Moriond4} & 5.0 fb$^{-1}+12.1$ fb$^{-1}$ \\\hline
$h\to\tau\tau$&$0.8\pm 0.7$~\cite{ATLAS_Moriond4}&4.6 fb$^{-1}+13$ fb$^{-1}$& $1.1\pm 0.4$~\cite{CMS_Moriond5} & 4.9 fb$^{-1}+19.4$ fb$^{-1}$ \\\hline
\end{tabular} 
\caption{Signal strengths of the five processes measured at ATLAS and CMS.}
\label{data}
\end{center}
\end{table}

So far, the Higgs boson search at the LHC has been done in the following five processes; 
$pp\to \gamma\gamma$, $pp\to ZZ^*$, $pp\to WW^*$, $pp\to \tau\tau$, $q\bar{q}'\to Vh\to Vb\bar{b}$, where $pp\to h$ and $q\bar{q}'\to Vh$ indicate the inclusive Higgs boson production and the vector boson associate production, respectively.  The signal strength for each of the channels is defined as
\begin{align}
\mu_X^{\text{Ref}} \equiv \frac{\sigma_{h}^{\text{Ref}}\times \text{BR}(h\to X)^{\text{Ref}}}{\sigma_{h}^{\text{SM}}\times 
\text{BR}(h\to X)^{\text{SM}}}, 
\end{align}
where $\sigma_{h}^{\text{Ref}}$ [$\sigma_{h}^{\text{SM}}$] and $\text{BR}(h\to X)^{\text{Ref}}$ [$\text{BR}(h\to X)^{\text{SM}}$] are the reference value [SM prediction] of the Higgs production cross section and that of the branching fraction of the $h\to X$ decay, respectively.  Experimental data of $\mu_X$ are listed in TABLE~\ref{data}.
The average signal strengths of the ATLAS and CMS data can be calculated using 
\begin{align}
\mu_{X}^{\text{exp}} \equiv \frac{\int\mathcal{L}^{\text{ATLAS}}\times \mu_{X}^{\text{ATLAS}}+ \int\mathcal{L}^{\text{CMS}}\times\mu_{X}^{\text{CMS}}
}{\int\mathcal{L}^{\text{ATLAS}}+\int\mathcal{L}^{\text{CMS}}},
\end{align}
where $\int \mathcal{L}^\text{ATLAS}$ ($\int \mathcal{L}^\text{CMS}$) is the integrated luminosity of the ATLAS (CMS) Collaboration.  We then obtain the average signal strengths as
\begin{align}
&\mu_{\gamma\gamma}^{\text{exp}} =1.22\pm 0.31,\quad 
\mu_{ZZ}^{\text{exp}} =1.21\pm 0.35,\quad 
\mu_{WW}^{\text{exp}} =0.89\pm 0.27,\notag\\
&\mu_{b\bar{b}}^{\text{exp}} =0.44\pm 0.87,\quad 
\mu_{\tau\tau}^{\text{exp}} =0.97\pm 0.55,   \label{ave}
\end{align}
where, to be conservative, the standard deviation of each signal strength is derived by using the larger error when the error bar is asymmetric. 
For $\mu_{\gamma\gamma}^{\text{exp}}$, we used the experimental value based on the MVA method.
With the input of Eq.~(\ref{ave}), a $\chi^2$ value can be calculated for each reference value as
\begin{align}
\chi^2 = \sum_X \left(\frac{\mu_X^{\text{exp}}-\mu_X^{\text{Ref}}}{\Delta \mu_X^{\text{exp}}}\right)^2, \label{chisq}
\end{align}
where $\Delta \mu_X^{\text{exp}}$ is the one standard deviation of $\mu_X^{\text{exp}}$. 

\begin{table}[t]
\begin{center}
\begin{tabular}{|l||c|c|}\hline 
Model& $\chi^2_{\text{min}}$&$\tan\beta$ \\\hline\hline
SM& 1.45 & - \\\hline
Type-I&  1.47  & 3.1 \\\hline
Type-II& 1.21  & 4.3 \\\hline
Type-X&  1.45  & 2.3 \\\hline
Type-Y&  1.24  & 4.5  \\\hline
\end{tabular} 
\caption{Minimal $\chi^2$ value ($\chi_{\text{min}}^2$) for the SM and the THDM's with $m_\Phi=M=300$ GeV and $\delta=10^{-4}$. }
\label{chisq_min}
\end{center}
\end{table}

We evaluate the $\chi^2$ values for the four scenarios (Type-I, Type-II, Type-X and 
Type-Y) defined in the previous section.  
In addition to the VEV $v = 246$ GeV and the mass of the SM-like Higgs boson $m_h = 126$ GeV, 
there are six free parameters in the Higgs sector: the three masses for the extra Higgs bosons $m_H$, $m_A$ and $m_{H^+}$, the two mixing angles $\alpha$ and $\beta$, and the dimensionful parameter $M^2$ representing the scale of softly broken $Z_2$ symmetry. 
In the following discussion, we consider the case where the masses of the extra Higgs bosons are taken to be degenerate, $m_\Phi\equiv m_H=m_A=m_{H^+}$, for simplicity and to avoid additional contributions to the electroweak $\rho$ parameter. 

In addition a large difference between $m_{\Phi}$ and $M$ means a large value of Higgs quartic couplings, 
it can be excluded by the constraints from the perturbative unitarity and also the vacuum stability, 
so that we take $M$ to be the same value as $m_{\Phi}$. 
The top Yukawa coupling is constrained by perturbativity to be $|y_t|^2<4\pi$ with $y_t= \sqrt{2}m_t/(v\sin\beta)$.  This gives the lower bound of $\tan\beta\gtrsim 0.3$.

We here summarize constraints from $B$ physics studies.  First, the $B\to X_s\gamma$ data demand that the mass of the charged Higgs boson, $m_{H^{+}}$, to be larger than 295 GeV at 95\% CL~\cite{Misiak} in the Type-II and Type-Y THDM's when $\tan\beta\gtrsim 2$.  When $\tan\beta\lesssim 2$, $m_{H^{+}}\lesssim 300$ GeV is excluded at 95\% CL in all types of THDM's~\cite{Stal}.  A similar but slightly weaker bound for $\tan\beta$ with $m_{H^{+}}\lesssim 300$ GeV has also been given by the $R_b$ data of the $Z\to b\bar{b}$ decay~\cite{Stal}. 
From the $B_u\to \tau\nu$, $B\to D\tau\nu$ and $K\to \mu\nu$ data, the Type-II THDM with $\tan\beta\gtrsim 30$ is excluded at 95\% CL\footnote{We note in passing that recently the BaBar Collaboration reported data on the ratios $\text{BR}(B \to D^{*} \tau \nu) / \text{BR}(B \to D^{*} \ell \nu)$ and $\text{BR}(B \to D \tau \nu) / \text{BR}(B \to D \ell \nu)$ ($\ell = e, \mu$) that deviate from the SM expectations by $2.7\sigma$ and $2.0\sigma$, respectively, and their combined deviation is $3.4\sigma$~\cite{Babar}.  These cannot be accommodated by simple versions of the THDM either because they favor different regions of $\tan\beta / m_{H^+}$.  A more conclusive result about this still awaits the corresponding analysis from the Belle Collaboration.} for $m_{H^{+}}= 300$ GeV~\cite{Stal,Haisch}.  In accord with the above-mentioned constraints, we will take $m_{\Phi}=300$ GeV and $\tan\beta\geq 2$ in the following numerical analyses.

We introduce a parameter $\delta\equiv 1-\sin(\beta-\alpha)$ to describe the deviation in the 
$hVV$ couplings from the corresponding SM values. 
Therefore, the mixing angle $\alpha$ is determined for a given pair of $\delta$ and $\tan\beta$. 
The ratios of the Higgs boson couplings $c_{hVV} \equiv g_{hVV}/g_{hVV}^{\text{SM}}$ and $c_{hff} \equiv g_{hff}/g_{hff}^{\text{SM}}$ are then
\begin{align}
c_{hVV}&=1-\delta, \qquad \mbox{for } V=W\text{ and }Z, \label{hvv} \\
c_{hff}&=1+\xi_f \sqrt{2\delta}+\mathcal{O}(\delta),
\qquad \mbox{for }~f=t,~b\text{ and }\tau,  \label{hff}
\end{align}
where the factor $\xi_f$ ($\xi_t=\xi_u$, $\xi_b=\xi_d$ and $\xi_\tau=\xi_e$) is listed in TABLE~\ref{Tab:type}.

\begin{figure}[t]
\begin{center}
\includegraphics[width=75mm]{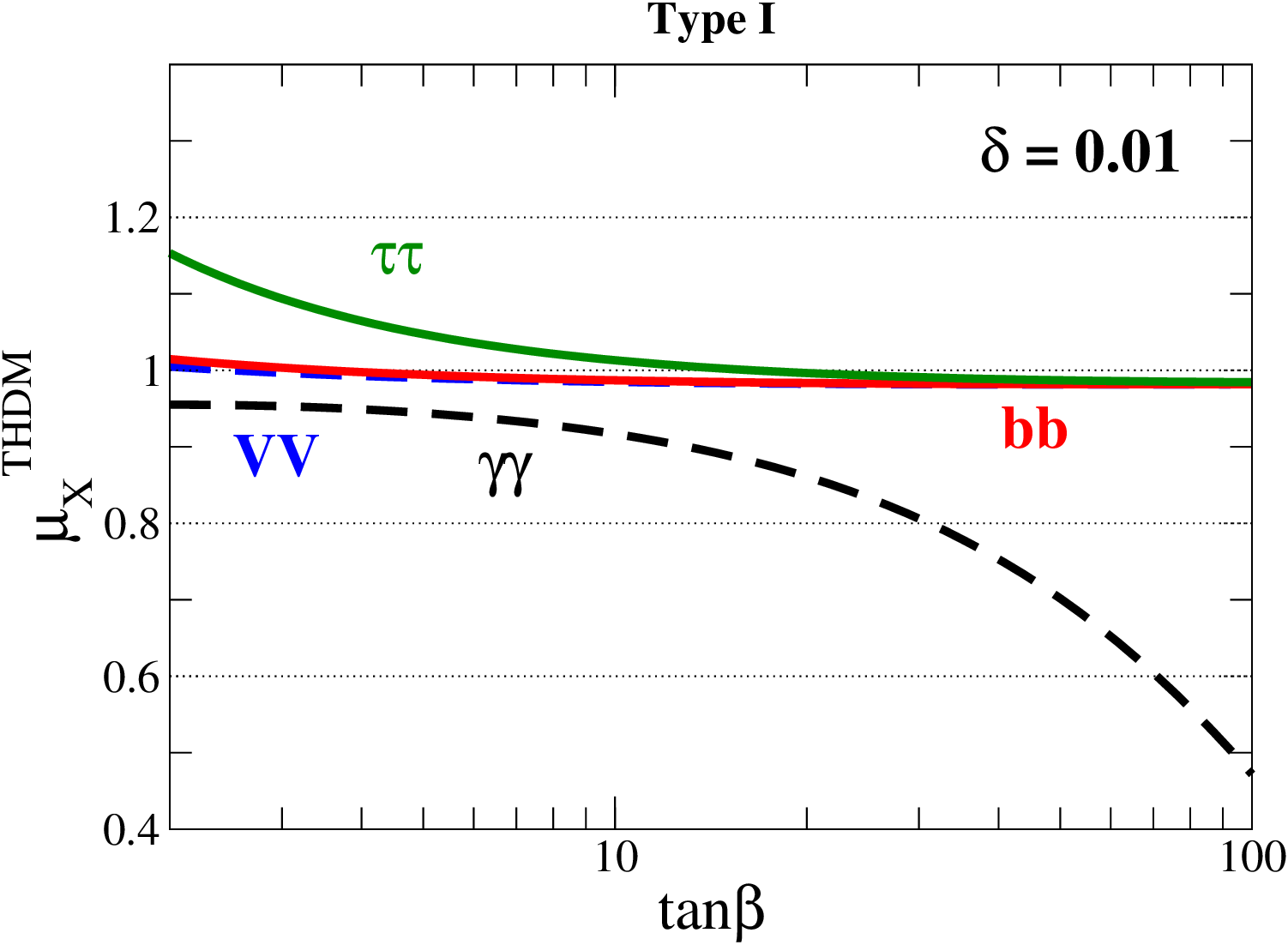}\hspace{5mm}
\includegraphics[width=75mm]{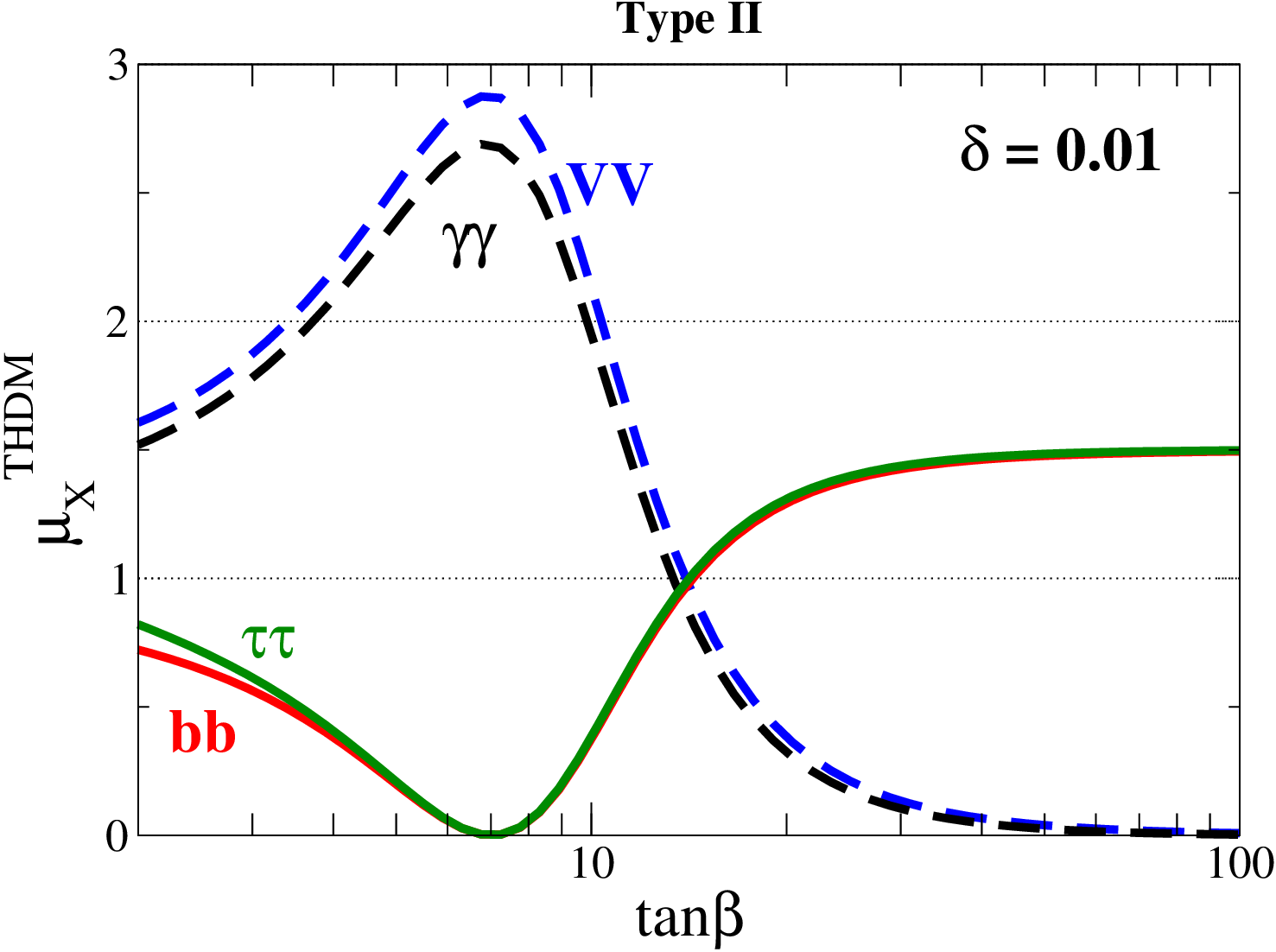}\\\vspace{5mm}
\includegraphics[width=75mm]{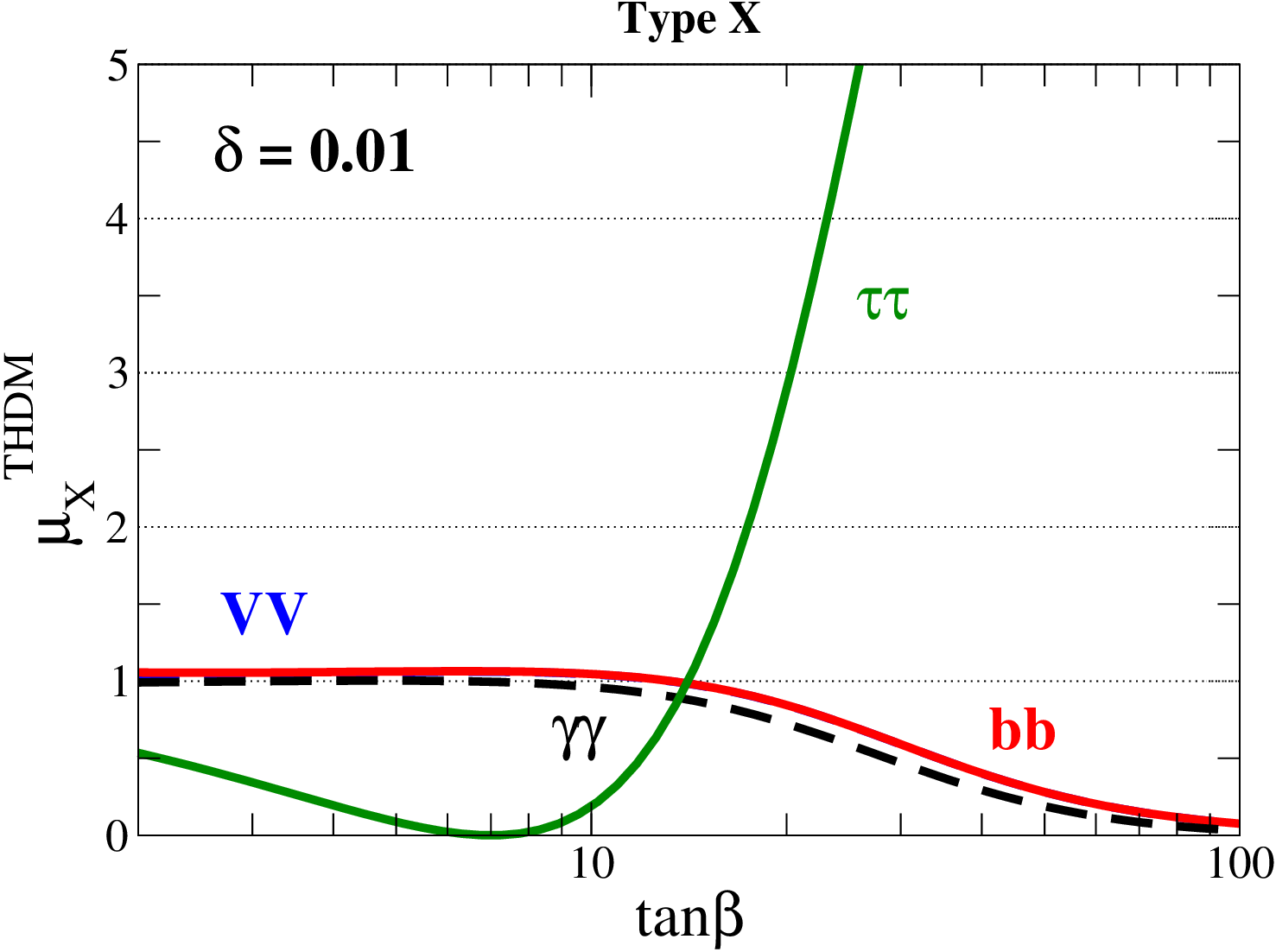}\hspace{5mm}
\includegraphics[width=75mm]{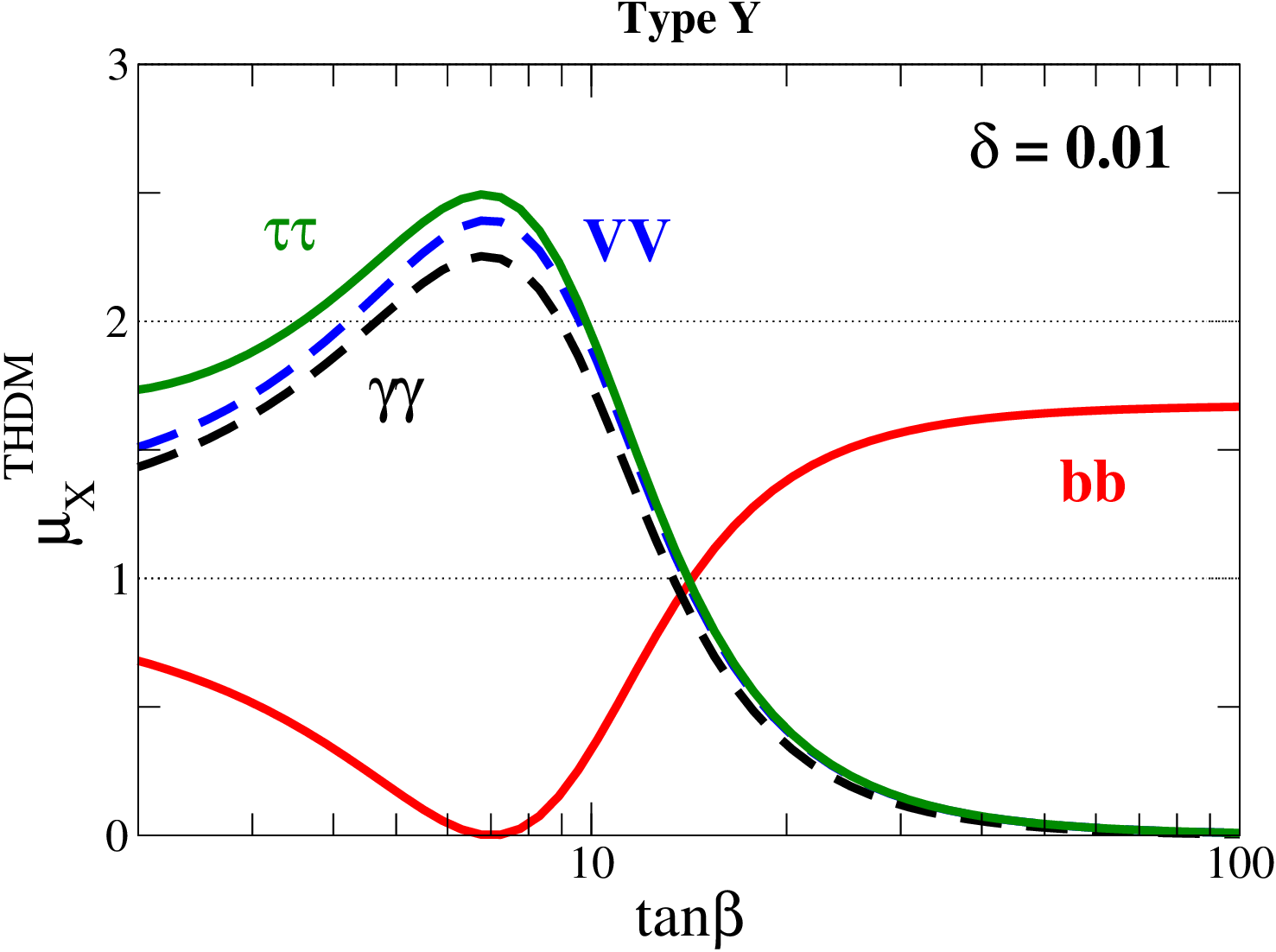}
\caption{$\mu_X^{\text{THDM}}$ as a function of $\tan\beta$ for the Type-I (upper left), Type-II (upper right), Type-X (lower left) and Type-Y (lower-right) THDM, with $\delta = 10^{-2}$ and $m_\Phi=M=300$ GeV. }
\label{mu_4types}
\end{center}
\end{figure}

\begin{figure}[t]
\begin{center}
\includegraphics[width=75mm]{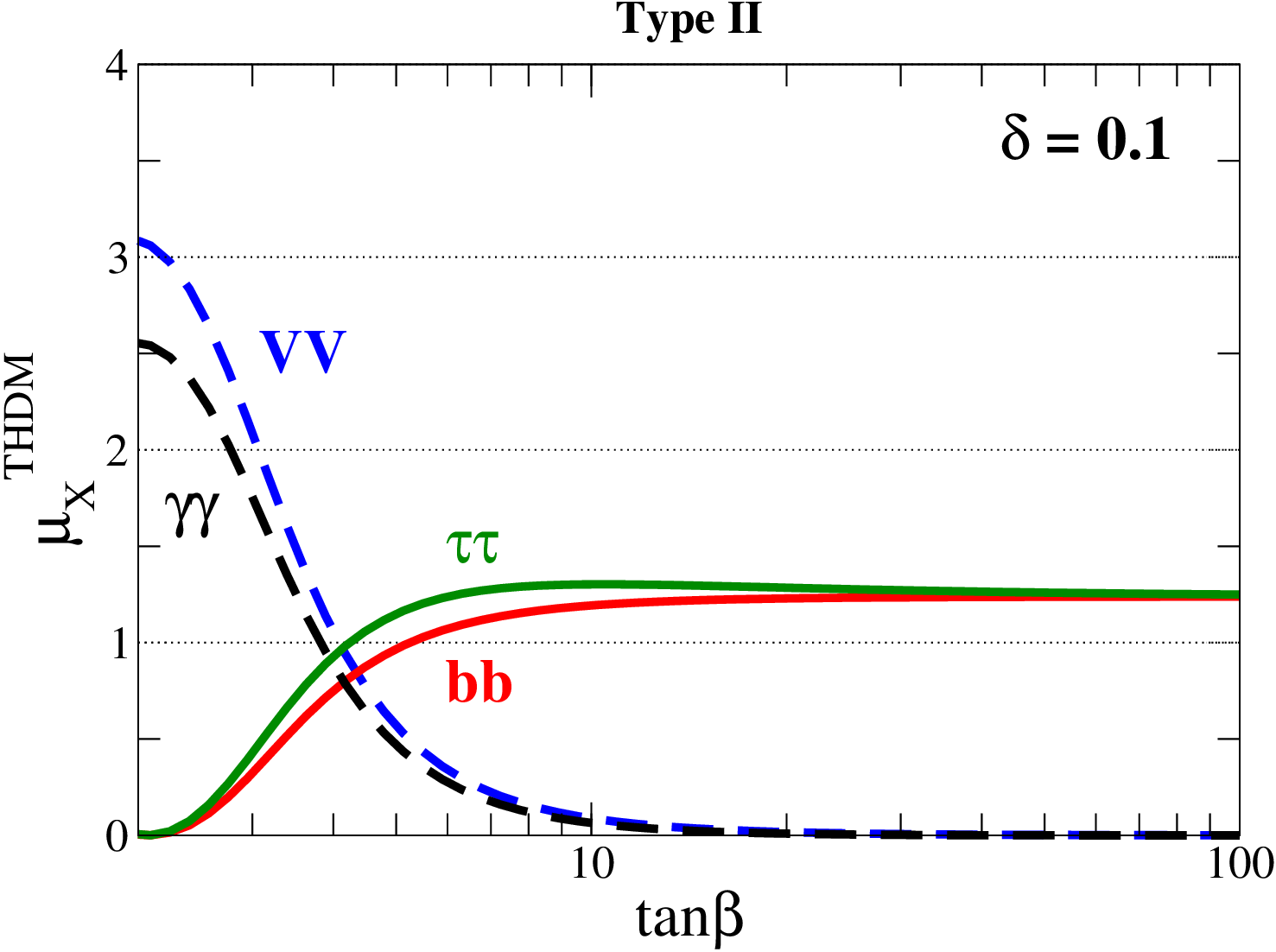}\hspace{5mm}
\includegraphics[width=75mm]{mu_del001_2.eps}\\\vspace{5mm}
\includegraphics[width=75mm]{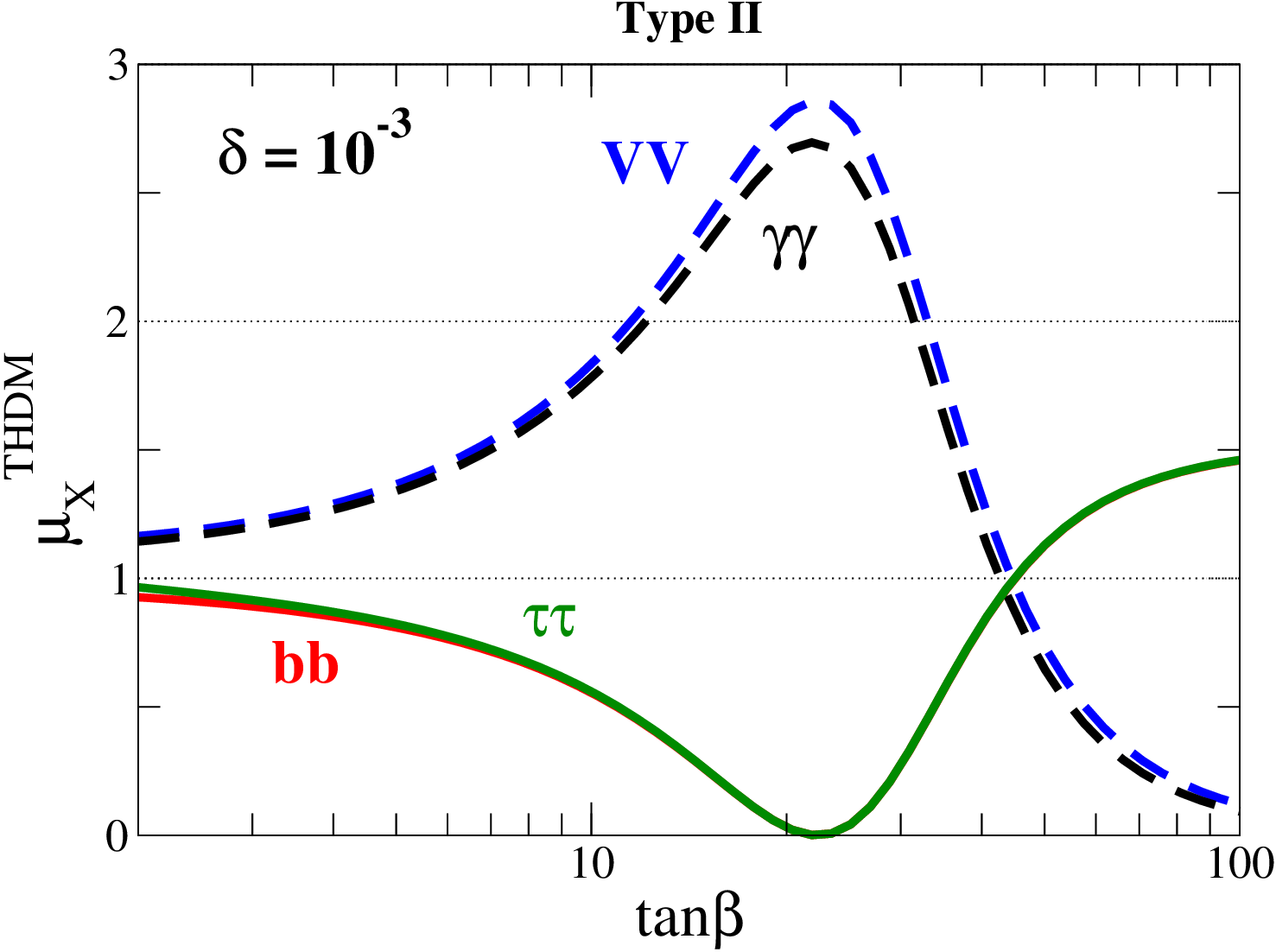}\hspace{5mm}
\includegraphics[width=75mm]{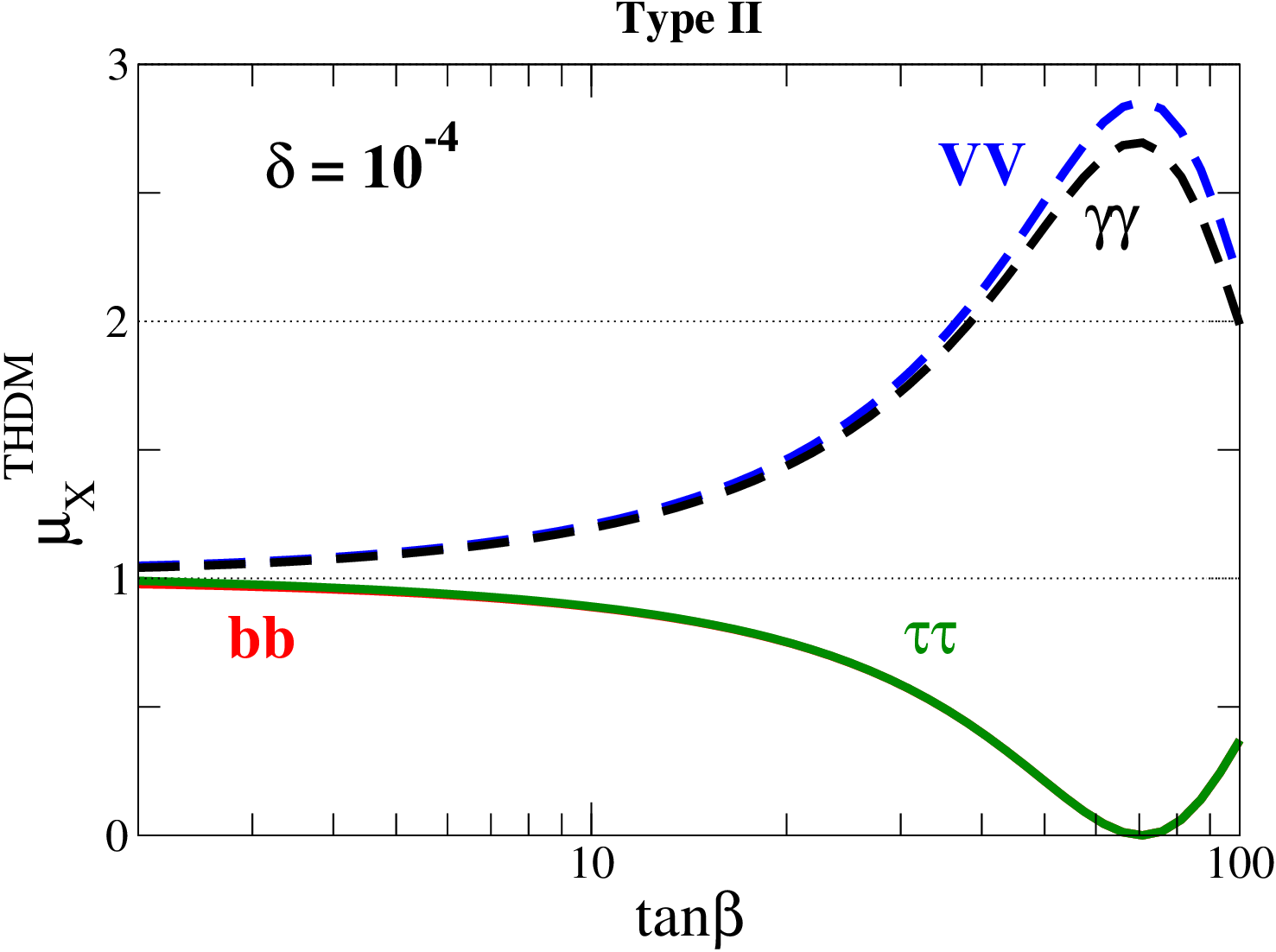}
\caption{$\mu_X^{\text{THDM}}$ as a function of $\tan\beta$ for the Type-II THDM with $m_\Phi=M=300$ GeV.  
The upper-left, upper-right, lower-left, and lower-right panels show 
the results for $\delta=0.1$, $10^{-2}$, $10^{-3}$ and $10^{-4}$, respectively. }
\label{mu_typeII}
\end{center}
\end{figure}

In TABLE~\ref{chisq_min}, the minimal value of $\chi^2$ ($\chi_{\text{min}}^2$) and the corresponding $\tan\beta$ value are listed for the SM and the THDM's with $\delta=10^{-4}$ and $m_\Phi=M=300$ GeV. 
The smallest value of $\chi_{\text{min}}^2$ among these models is obtained in the Type-II THDM with $\tan\beta=4.3$.  
Except for the Type-I THDM, the value of $\chi_{\text{min}}^2$ stays almost the same for $\delta<10^{-4}$.  But the value of $\tan\beta$ at the $\chi^2$ minimum gets larger than that for $\delta=10^{-4}$. 
In the Type-I THDM, the value of $\chi_{\text{min}}^2$ approaches $3.26$, still larger than the SM value, in the limit of $\delta\to 0$ with $\tan\beta\simeq 4.3$. 
This is due to a destructive $H^\pm$ loop contribution to the decay rate of $h\to \gamma\gamma$ with the $W$ boson loop contribution (see Eq.~(\ref{hchch})).

In Fig.~\ref{mu_4types}, values for $\mu_X^{\text{THDM}}$ are plotted as a function of $\tan\beta$ in the Type-I (upper-left), Type-II (upper-right), Type-X (lower-left) and Type-Y (lower-right) THDM's with $\delta=10^{-2}$. 
It is seen that in the Type-II THDM, both $\mu_{bb}^{\text{THDM}}$ and $\mu_{\tau\tau}^{\text{THDM}}$ are around $1.5$ while all the others approach 0 in the large $\tan\beta$ region.  This is
because both bottom and tau Yukawa couplings are enhanced by the same factor $1-\tan\beta\sqrt{2\delta}$, while the top Yukawa and gauge couplings are almost the same as their SM values. 
Thus, the Higgs production cross section is barely changed from the SM prediction.  Yet the branching fractions of $h\to \tau\tau$ and $h\to bb$ modes are as large as about 10\% and 90\%, respectively, roughly $1.5$ times larger than the SM values. 
On the other hand, in the Type-X (Type-Y) THDM, only the tau (bottom) Yukawa coupling can be enhanced by 
the factor of $1-\tan\beta\sqrt{2\delta}$, so that only $\mu_{\tau\tau}^{\text{THDM}}$ ($\mu_{bb}^{\text{THDM}}$) 
is larger than 1 in the large $\tan\beta$ region. 
The asymptotic values of $\mu_{\tau\tau}^{\text{THDM}}$ ($\mu_{bb}^{\text{THDM}}$) are 
about $14$ ($1.7$) in the Type-X (Type-Y) THDM, respectively, in the large $\tan\beta$ limit. 

We note that all the $\mu_X^{\text{THDM}}$ values are around 1 at $\tan\beta\simeq 15$ 
where $c_{hbb}$ and $c_{h\tau\tau}$ (Type-II), $c_{h\tau\tau}$ (Type-X) and $c_{hbb}$ (Type-Y) have an opposite sign to their corresponding SM couplings, 
while the other $c_{hff}$'s and $c_{hVV}$ are almost the same as the SM values.
In the Type-I THDM, all the Yukawa couplings are modified by the same factor, and they are close to the SM values in the large $\tan\beta$ region so that all the $\mu_{X}^{\text{THDM}}$ values approach 1 except for $\mu_{\gamma\gamma}^{\text{THDM}}$. 
Only the value of $\mu_{\gamma\gamma}^{\text{THDM}}$ becomes smaller than 1 for large $\tan\beta$  because of the destructive contribution of the charged Higgs boson loop~\footnote{
The same effect occurs in the other three models as well.  
But it is masked by the large branching fractions of $h\to f\bar{f}$ modes; {\it e.g.}, 
large branching fractions of $h\to b\bar{b}$ and $h\to \tau\tau$ in the Type-II THDM.}.

In Fig.~\ref{mu_typeII}, values of $\mu_X^{\text{THDM}}$ are displayed as a function of $\tan\beta$ in the Type-II THDM with $\delta =0.1$ (upper-left), $10^{-2}$ (upper-right), $10^{-3}$ (lower-left), and $10^{-4}$ (lower-left).  It is seen that in the cases with smaller values of $\delta$, 
the value of $\mu_{VV}^{\text{THDM}}$ ($\mu_{bb}^{\text{THDM}}$) gets quite close to that of $\mu_{\gamma\gamma}^{\text{THDM}}$ ($\mu_{\tau\tau}^{\text{THDM}}$), because the decay rates of $h\to b\bar{b}$ and $h\to \tau\tau$ can be changed by the same factor $c_{hbb}^2=c_{h\tau\tau}^2= (1-\tan\beta\sqrt{2\delta})^2$, and the other decay rates ($h\to \gamma\gamma$ and $h\to VV$) and the production cross sections (the gluon fusion and the gauge boson associated production) are almost the same as in the SM. 
We observe a peak for both $\mu_{\gamma\gamma}^{\text{THDM}}$ and $\mu_{VV}^{\text{THDM}}$, 
especially for the case with $\delta < 0.1$.  This is caused by vanishing bottom and tau Yukawa couplings.  The location of the peak can be determined by solving $c_{hbb}=c_{h\tau\tau}=0$, from which $\tan\beta \simeq 1/\sqrt{2\delta}$.

\begin{figure}[t]
\begin{center}
\includegraphics[width=75mm]{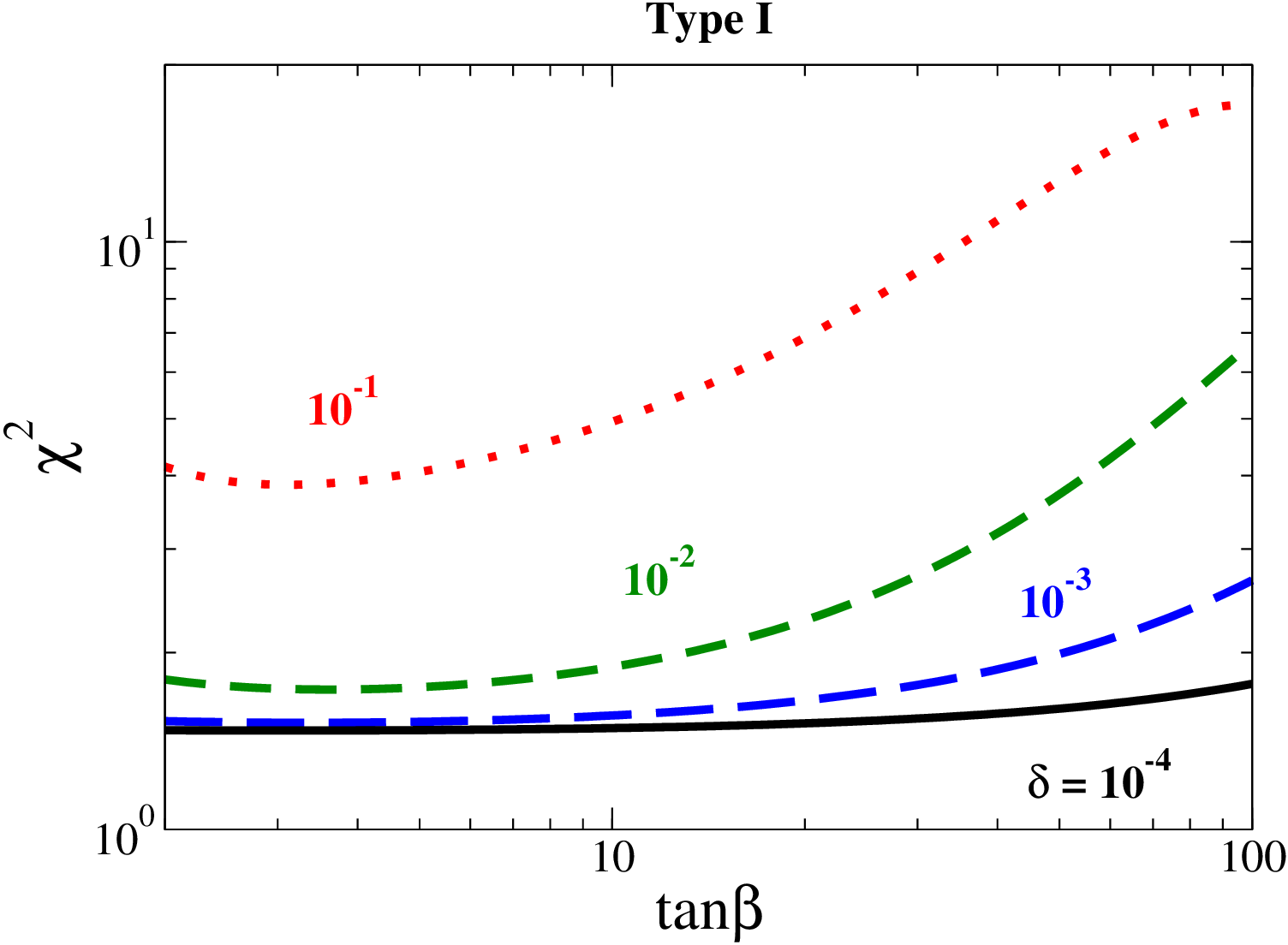}\hspace{5mm}
\includegraphics[width=75mm]{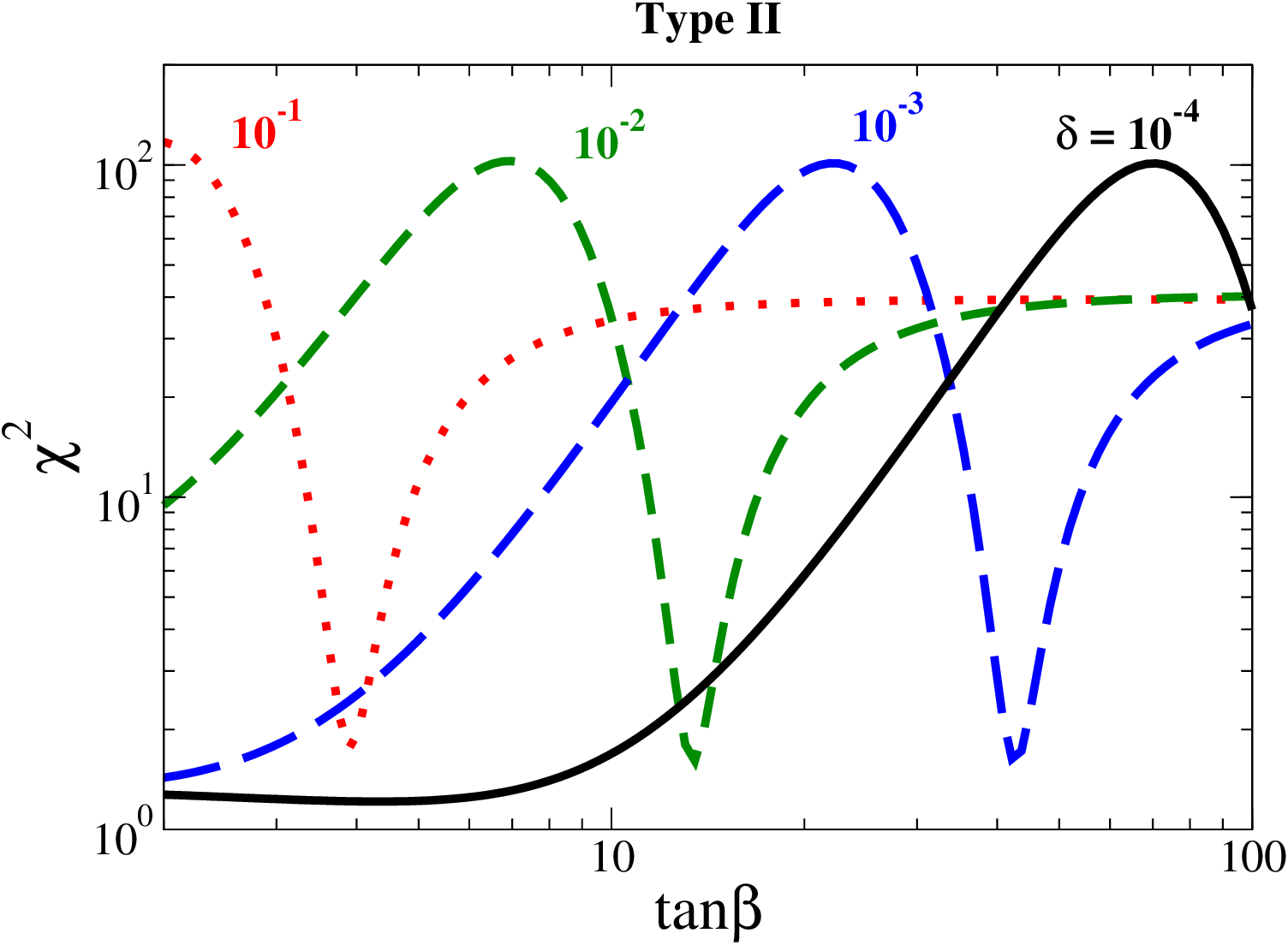}\\\vspace{5mm}
\includegraphics[width=75mm]{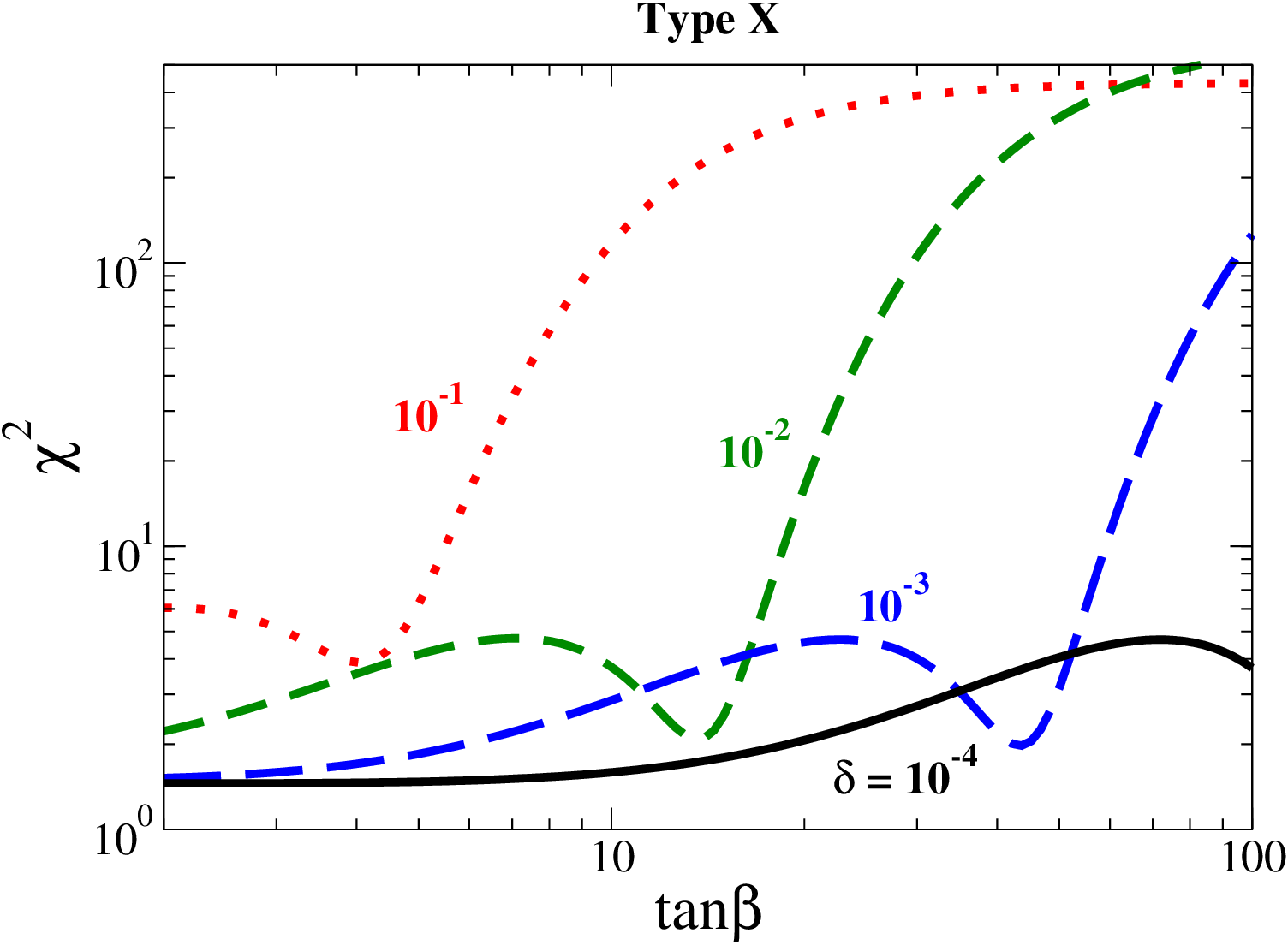}\hspace{5mm}
\includegraphics[width=75mm]{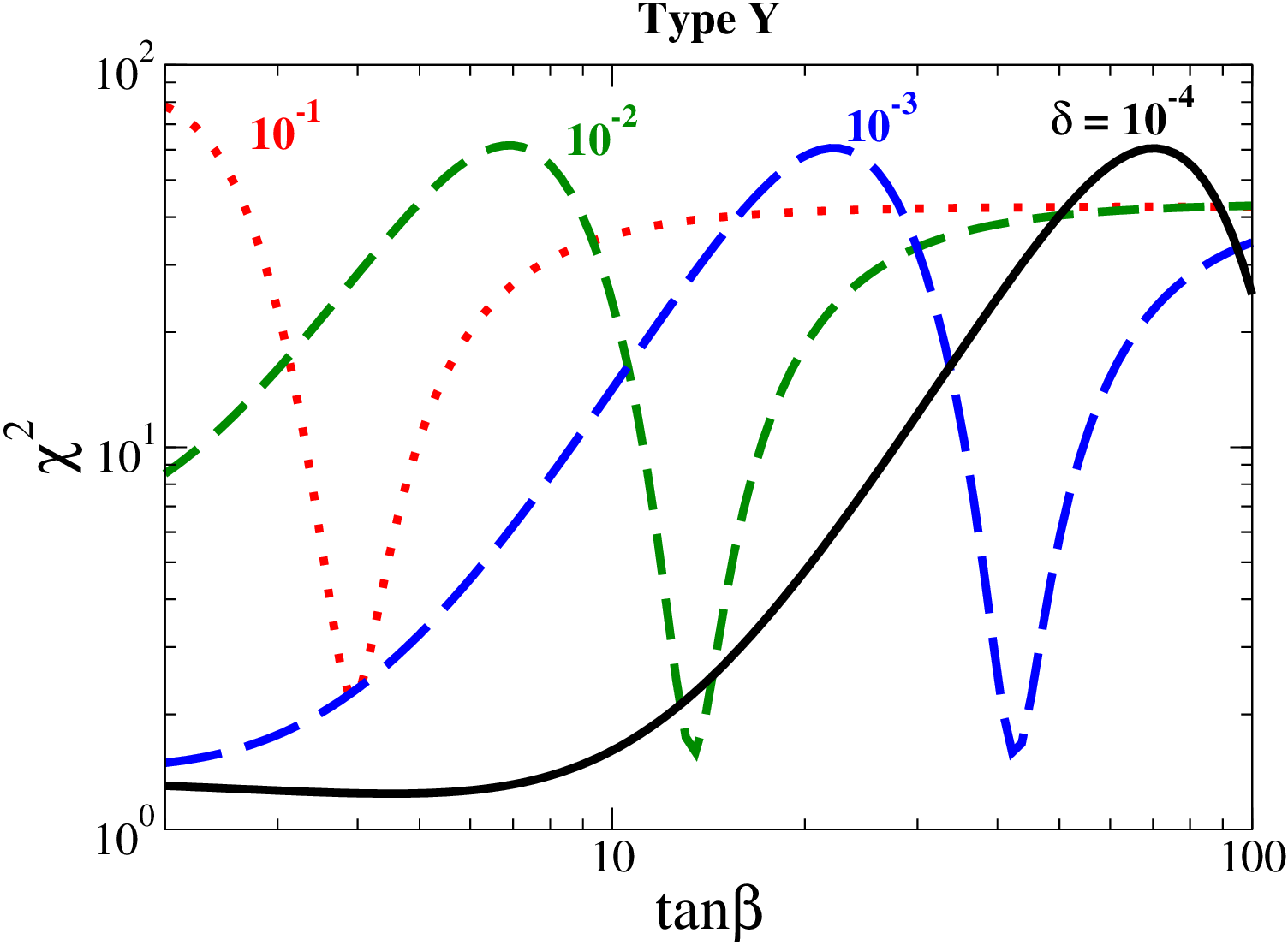}
\caption{$\chi^2$ as a function of $\tan\beta$ for several values of $\delta$ and $m_\Phi=M=300$ GeV.  The upper-left, upper-right, lower-left and lower-right panels show the results in the Type-I, Type-II, Type-X and Type-Y THDM, respectively. 
}
\label{chisq_2}
\end{center}
\end{figure}

In Fig.~\ref{chisq_2}, $\chi^2$ is plotted as a function of $\tan\beta$ for $\delta=0.1, 10^{-2}, 10^{-3}$ and $10^{-4}$ in the THDM's of Type-I (upper-left), Type-II (upper-right), Type-X (lower-left) and Type-Y (lower-right). 
In the Type-II and Type-Y THDM, there is a valley in the curve for each case of $\delta$ except for $\delta=10^{-4}$, where the bottom of the valley gives the minimal $\chi^2$ value. 
The value of $\tan\beta$ corresponding to the bottom of the valley is slightly smaller than the one for all the $\mu_{X}^{\text{THDM}}$ values to be unity, as seen in Fig.~\ref{mu_4types}. 
In the Type-X THDM, there is a similar valley in each value of $\delta$ except for $\delta=10^{-4}$, yet the bottom of that does not correspond to the global $\chi^2$ minimum. 
This is because the $h\to \gamma\gamma$ channel is not sufficiently enhanced in comparison with that in the Type-II and Type-Y THDM's (see the lower-left plot in Fig.~\ref{mu_4types}). 
In the Type-I THDM, there is no such a valley, and the minimum of $\chi^2$ is obtained at around $\tan\beta=3.1$ independent of the value of $\delta$.

\begin{figure}[t]
\begin{center}
\includegraphics[width=75mm]{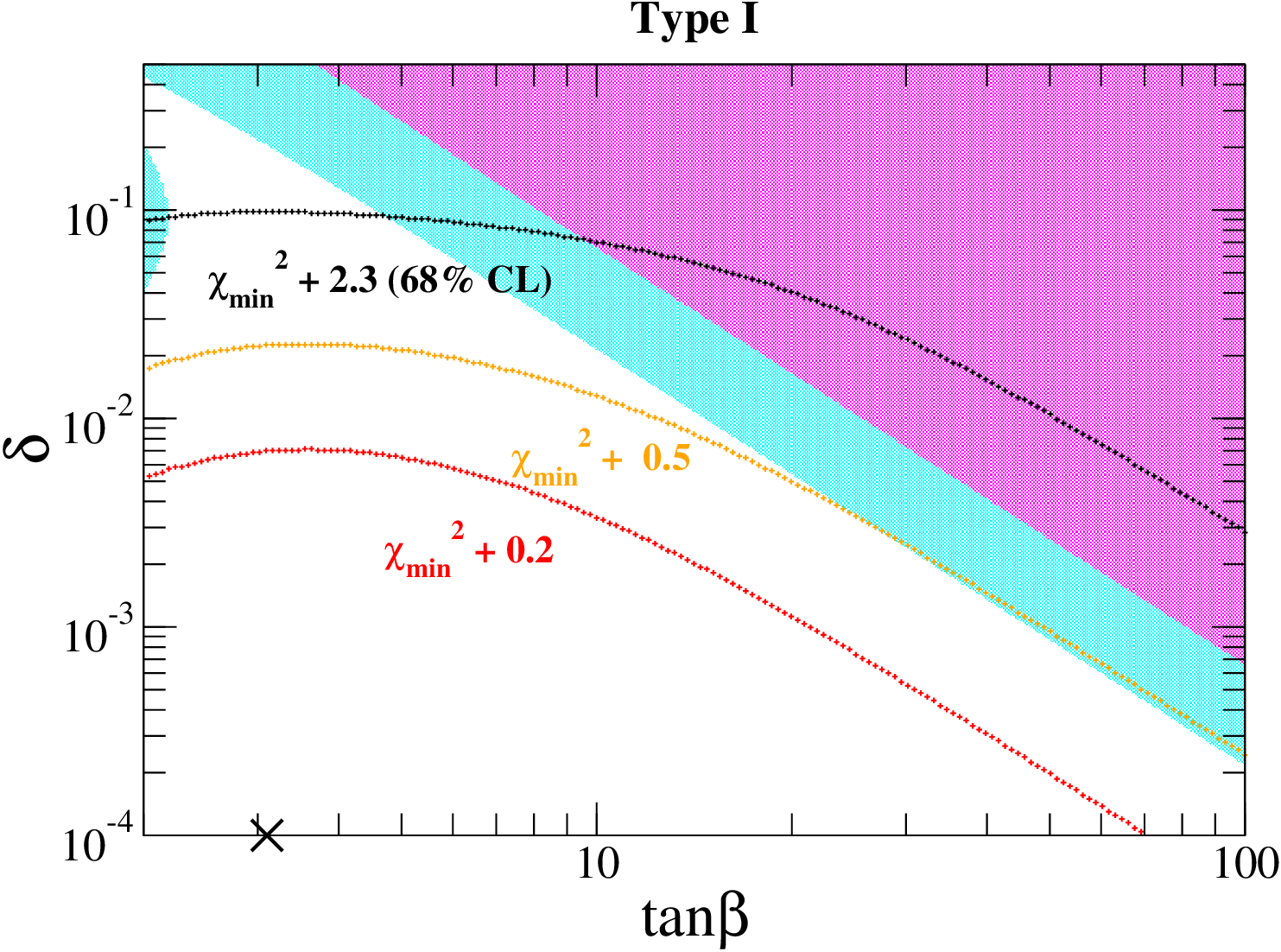}\hspace{5mm}
\includegraphics[width=75mm]{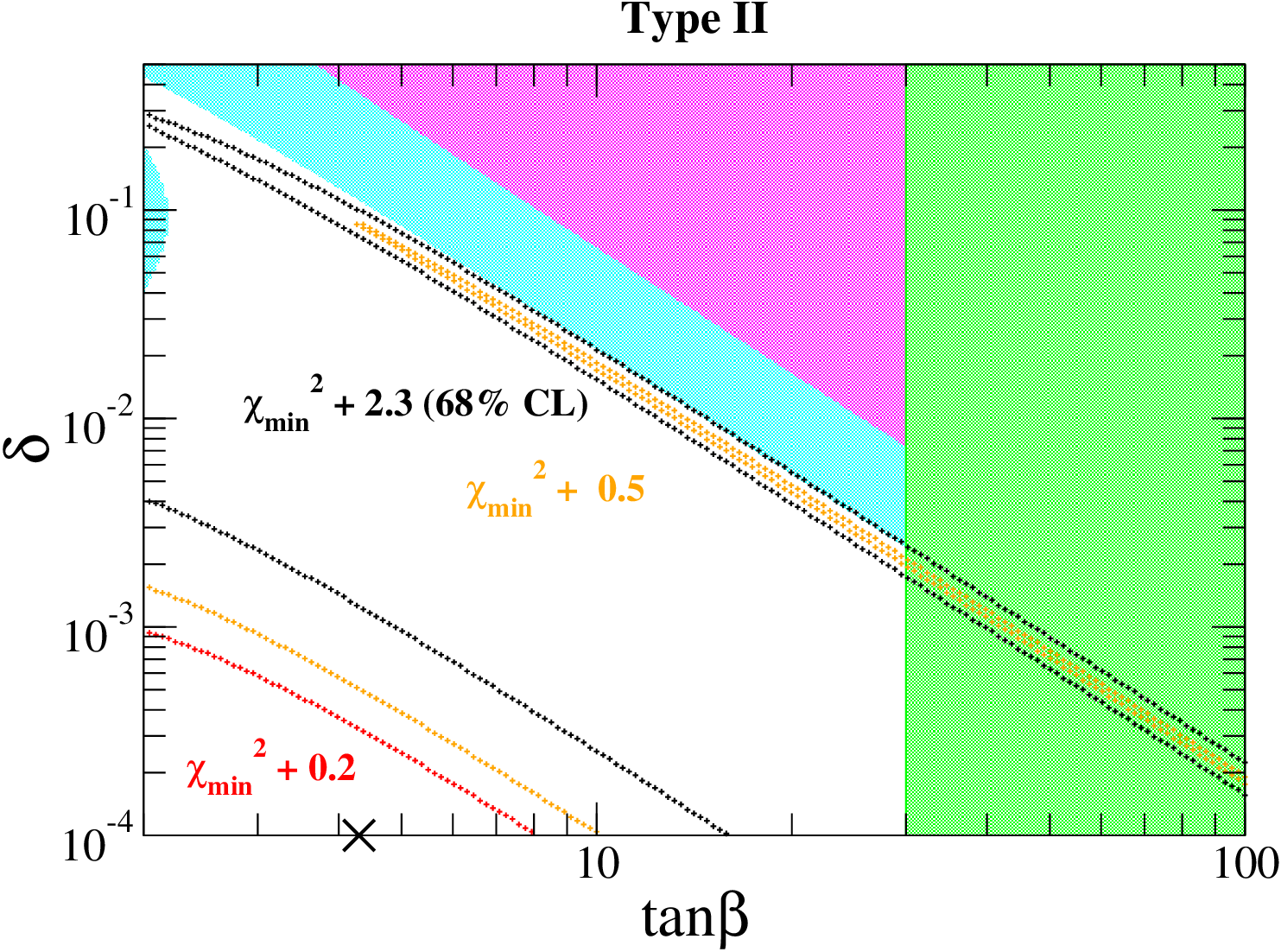}\\\vspace{5mm}
\includegraphics[width=75mm]{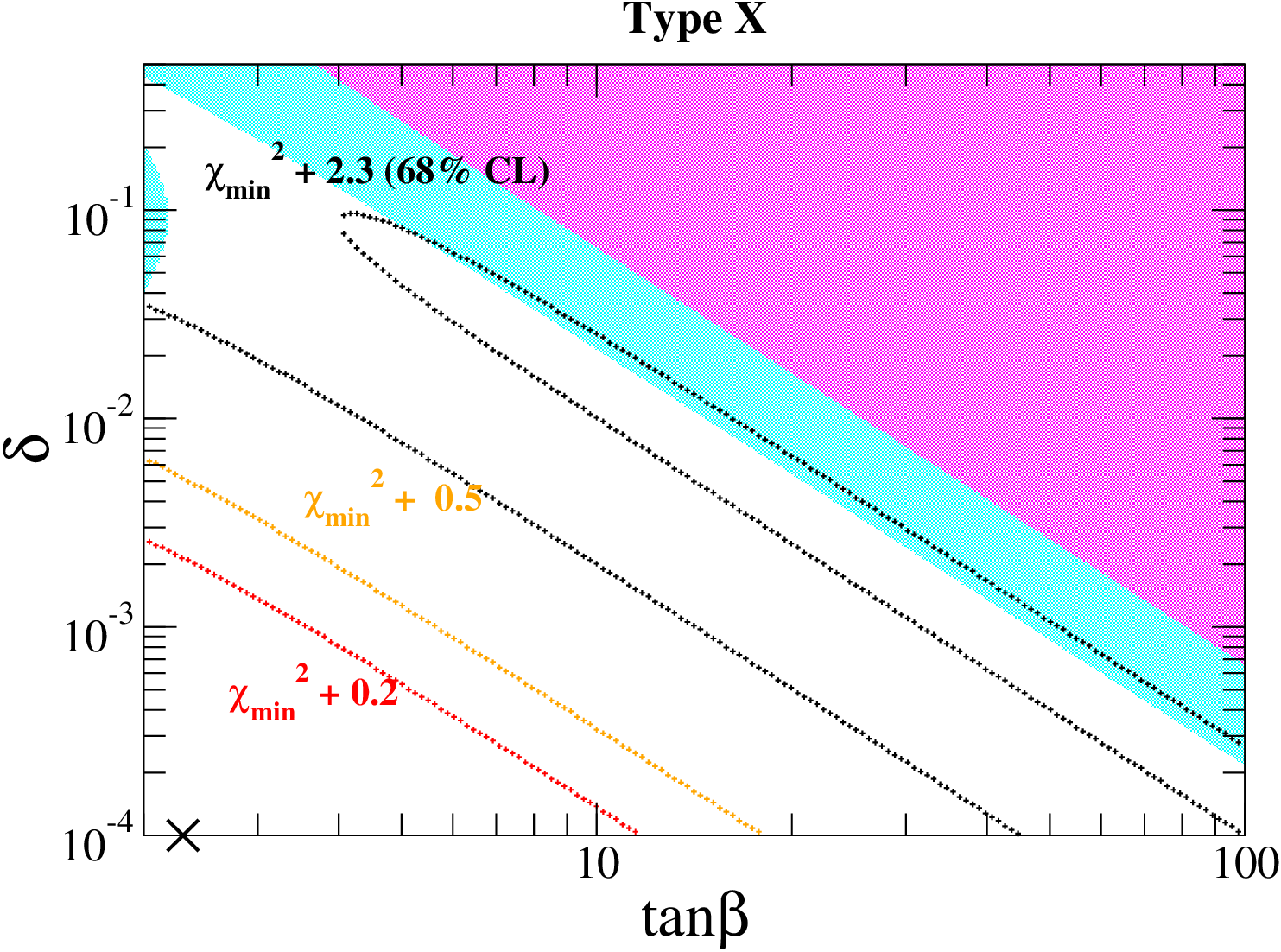}\hspace{5mm}
\includegraphics[width=75mm]{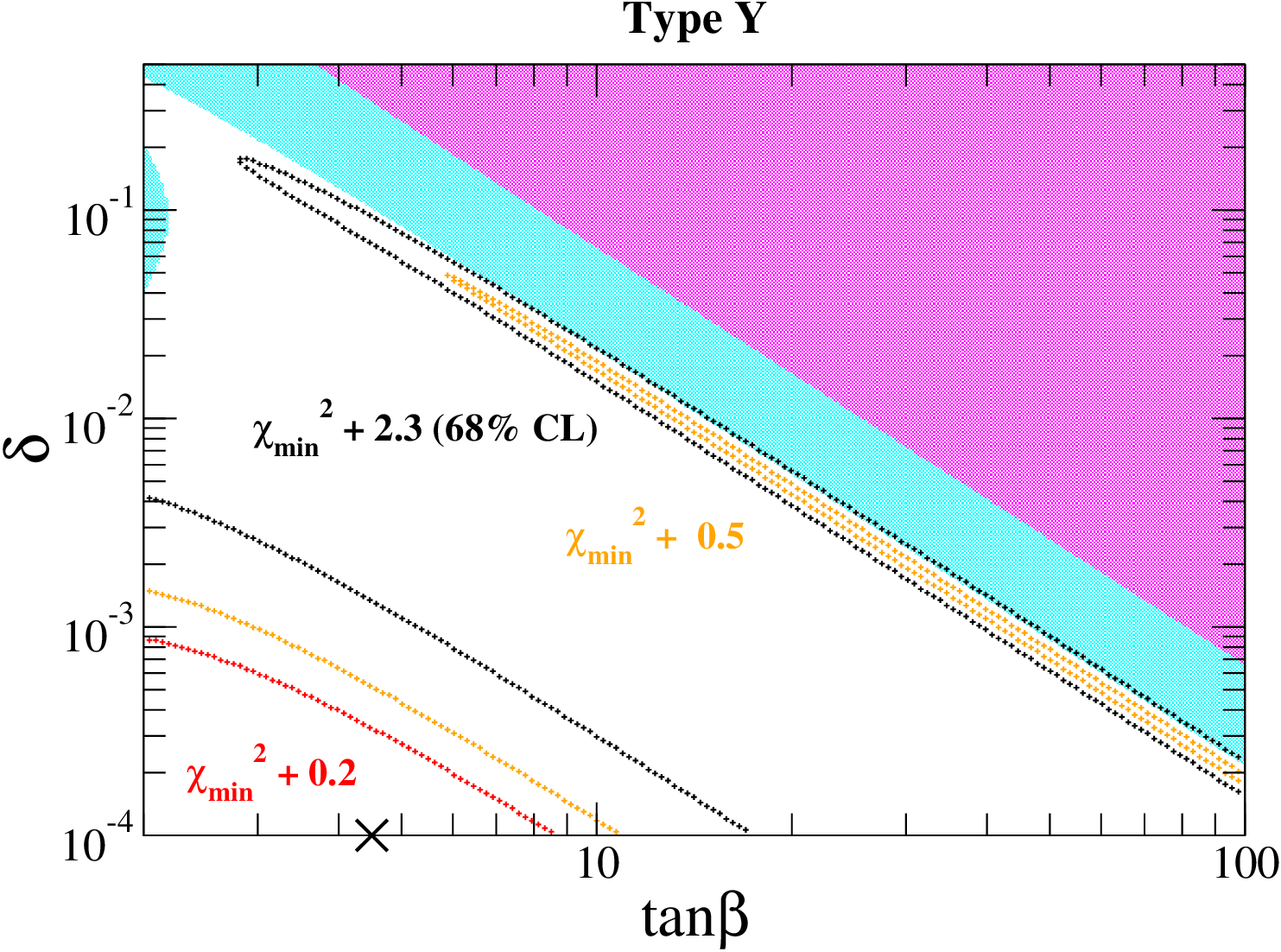}
\caption{Contour plots of $\chi^2$ values in the $\tan\beta$-$\delta$ plane with $m_\Phi=M=300$ GeV.  The upper-left, upper-right, lower-left and lower-right panels show the results for the models of Type-I, Type-II, Type-X and Type-Y, respectively. 
The coordinate marked by ${\bm\times}$ has the minimum $\chi^2$ value. 
The red, orange and black curves are respectively contours for $\chi_{\text{min}}^2+0.2$, 
$\chi_{\text{min}}^2+0.5$ and $\chi_{\text{min}}^2+2.3$ (68\% CL). 
The light blue (dark magenta) regions are excluded by the constraint of vacuum stability (perturbative unitarity).  The light green region in the upper-right panel is excluded by the $B$ physics studies mentioned in the main text.}
\label{contours}
\end{center}
\end{figure}

Fig.~\ref{contours} shows the contour plot for $\chi^2$ in the $\tan\beta$-$\delta$ plane 
in the Type-I (upper-left), Type-II (upper-right), Type-X (lower-left) and Type-Y (lower-right) THDM's.  The cross ${\bm\times}$ indicates the point that gives the minimal $\chi^2$ value, $\chi^2_{\text{min}}$.  The red, orange and black curves are respectively the contours corresponding to $\chi^2_{\text{min}}+0.2$, $\chi^2_{\text{min}}+0.5$ and $\chi^2_{\text{min}}+2.3$, the last case corresponding to $68\%$ CL.  The light blue (dark magenta) regions are excluded by the constraint of vacuum stability~\cite{vs_THDM} (perturbative unitarity~\cite{pv_THDM}).  
In the upper-right figure, the light green area shows the excluded region by the $B$ physics studies.   
It is seen that there is an isolated narrow region consistent with data at 68\% CL for Type-II, Type-X and Type-Y THDM's, and is not excluded by the vacuum stability and perturbative unitarity conditions.  The values of $\tan\beta$ and $\delta$ in this region correspond roughly to the bottom of the valleys in Fig.~\ref{chisq_2}.  

From the above analysis, we here conclude that Type-II THDM can best explain the current LHC data among all the THDM's with softly broken $Z_2$ symmetry.  This is because the branching fraction of the $h\to \gamma\gamma$ mode can be enhanced due to the suppressed decay rates of 
$h\to b\bar{b}$ and $h\to \tau\tau$ modes.  There are two regions where the data can be well reproduced in the model of Type-II.  One of them is the region with a very small $\delta$; {\it e.g.}, $\delta \alt 0.01$.  The other is the region satisfying $c_{hbb}\simeq c_{h\tau\tau}\simeq -1$,  which gives the relation $\delta \tan\beta\simeq \sqrt{2 \delta}$.  This is a good approximation especially for small values of $\delta$.

\section{Phenomenology of extra Higgs bosons \label{sec:pheno}}

In this section, we discuss the phenomenology of the extra Higgs bosons, {\it i.e.}, the heavier CP-even Higgs boson $H$, the CP-odd Higgs boson $A$ and the charged Higgs bosons $H^\pm$ in the parameter regions favored by the LHC data.  In the previous section, we found that the LHC data could be better explained in the Type-II THDM compared to the other THDM's.  Therefore, we focus on this scenario with $m_\Phi=M=300$ GeV as in the previous section. 
Furthermore, when $\delta$ is larger than $10^{-2}$, the favored parameter space on the $\tan\beta$-$\delta$ plane is restricted to a narrow band, so that the value of $\tan\beta$ can be approximately determined for each given value of $\delta$. 
Such larger deviations in the $hZZ$ and $hWW$ couplings ($\delta\gtrsim 10^{-2}$)
are expected to be measurable at future linear colliders such as the International Linear Collider~\cite{Peskin}.

The extra Higgs boson searches have been reported in Refs.~\cite{MSSM_charged_CMS,MSSM_charged_ATLAS,MSSM_neutral_CMS,MSSM_neutral_ATLAS} 
for the MSSM whose Higgs sector corresponds to that of the Type-II THDM with a supersymmetric relation among the Higgs parameters. 
So far, the search for the charged Higgs bosons has only been done in the $H^\pm\to \tau^\pm \nu$ and $H^\pm \to cs$ channels via the top quark decay $t\to H^+ b$. 
When the charged Higgs boson mass is larger than the top quark mass, no bound on its mass is currently available at the LHC~\cite{MSSM_charged_CMS,MSSM_charged_ATLAS}. 
On the other hand, the search for the extra neutral Higgs bosons $H$ and $A$ has been done in the 
gluon fusion process: $gg\to H/A \to \tau\tau/\mu\mu$ and the bottom quark associate processes: 
$gg\to b\bar{b}  H/A\to  b\bar{b}\tau\tau$. 
Current excluded regions in the $m_A$-$\tan\beta$ plane have been shown in Ref.~\cite{MSSM_neutral_CMS} by CMS and in Ref.~\cite{MSSM_neutral_ATLAS} by ATLAS. 
For example, if $m_A$ is taken to be 300 GeV, the upper bound for the value of $\tan\beta$ 
is given by $7.58$~\cite{MSSM_neutral_CMS}. 
We note that this bound for $\tan\beta$ can be modified in the (non-supersymmetric) 
Type-II THDM, for the production rates for $H$ and $A$ and also the decay branching ratios of $H/A\to \tau\tau$ can be different from those in the MSSM.

\begin{figure}[t]
\begin{center}
\includegraphics[width=75mm]{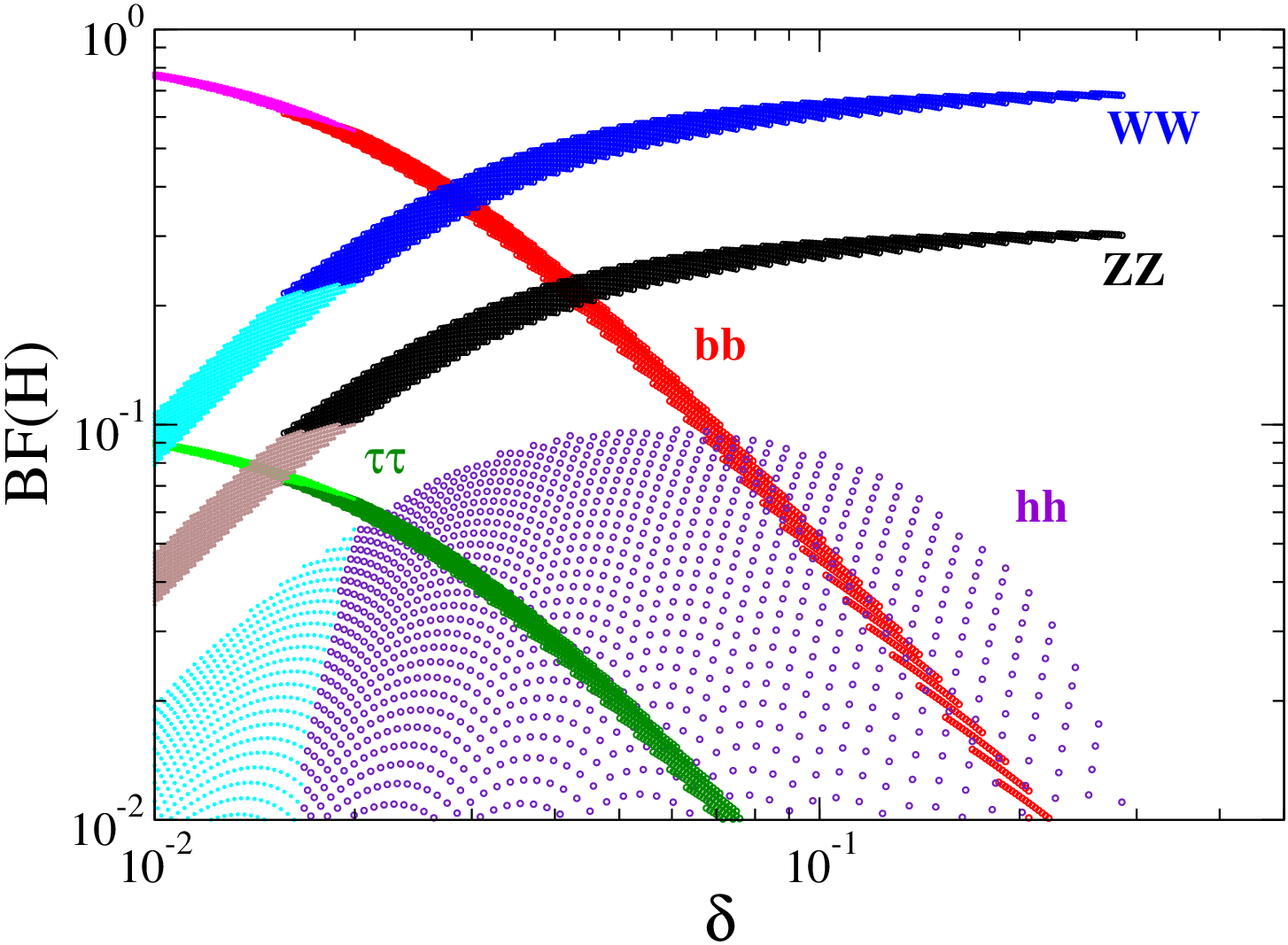}\hspace{5mm}
\includegraphics[width=75mm]{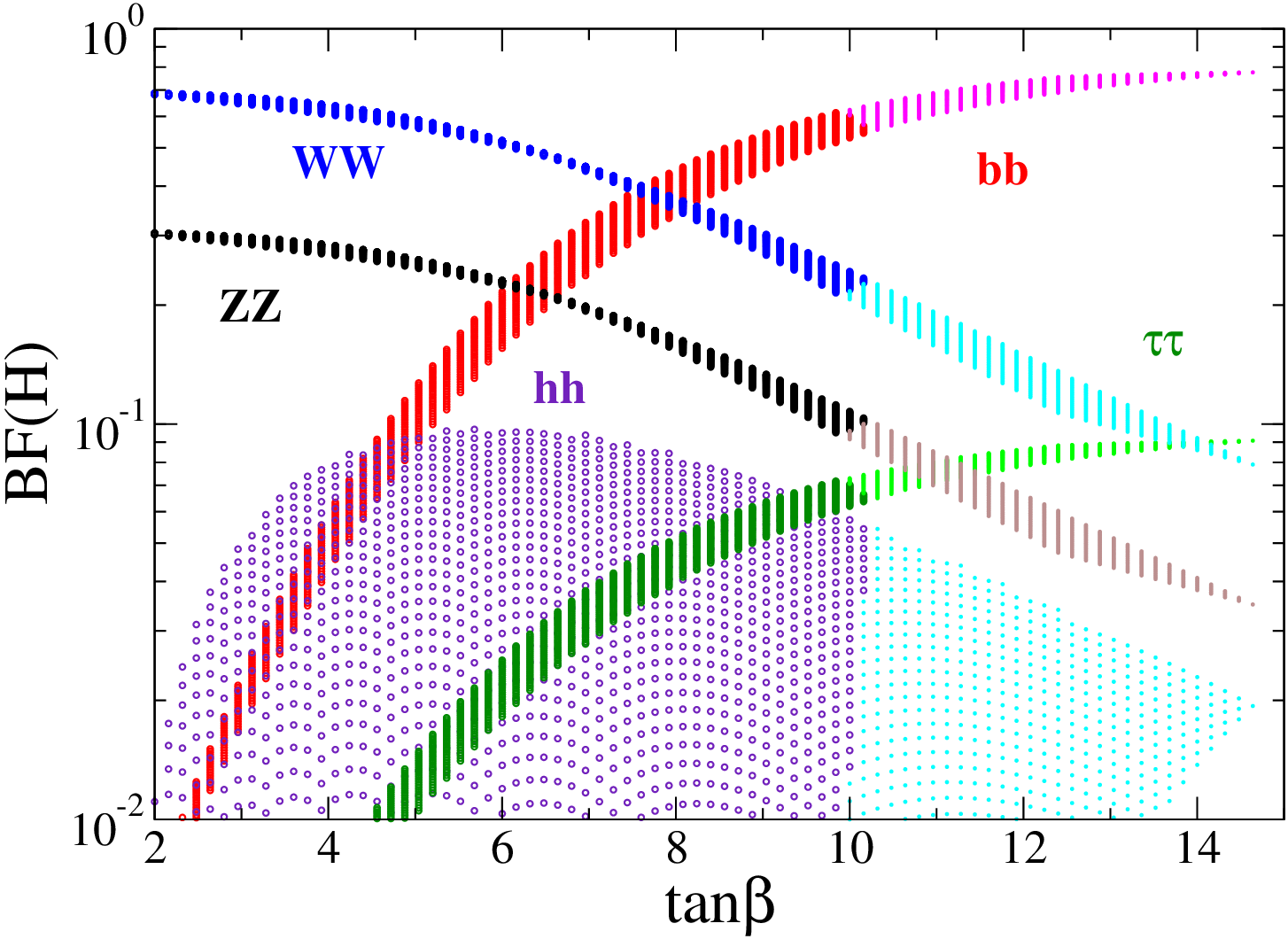}\\\vspace{5mm}
\caption{Decay branching fractions of $H$ as a function of $\delta$ (left panel) and $\tan\beta$ (right panel) for $m_\Phi =M=300$ GeV in the Type-II THDM.  The band for each curve reflects parameter uncertainty at the 68\% CL as determined by the Higgs data.  The light-colored parts are excluded by the LHC search data~\cite{MSSM_neutral_CMS}.}
\label{decay_H}
\end{center}
\end{figure}

\begin{figure}[t]
\begin{center}
\includegraphics[width=75mm]{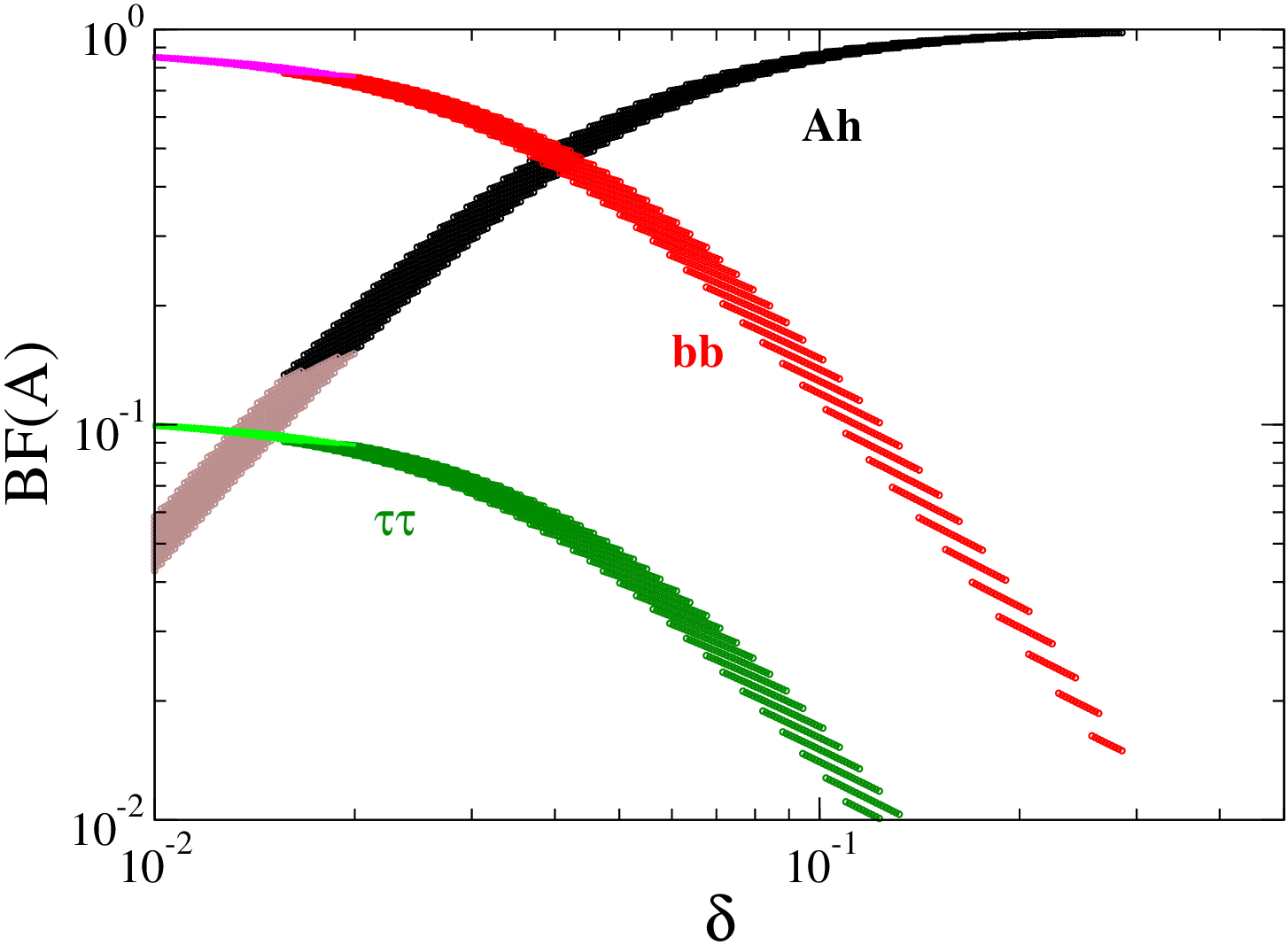}\hspace{5mm}
\includegraphics[width=75mm]{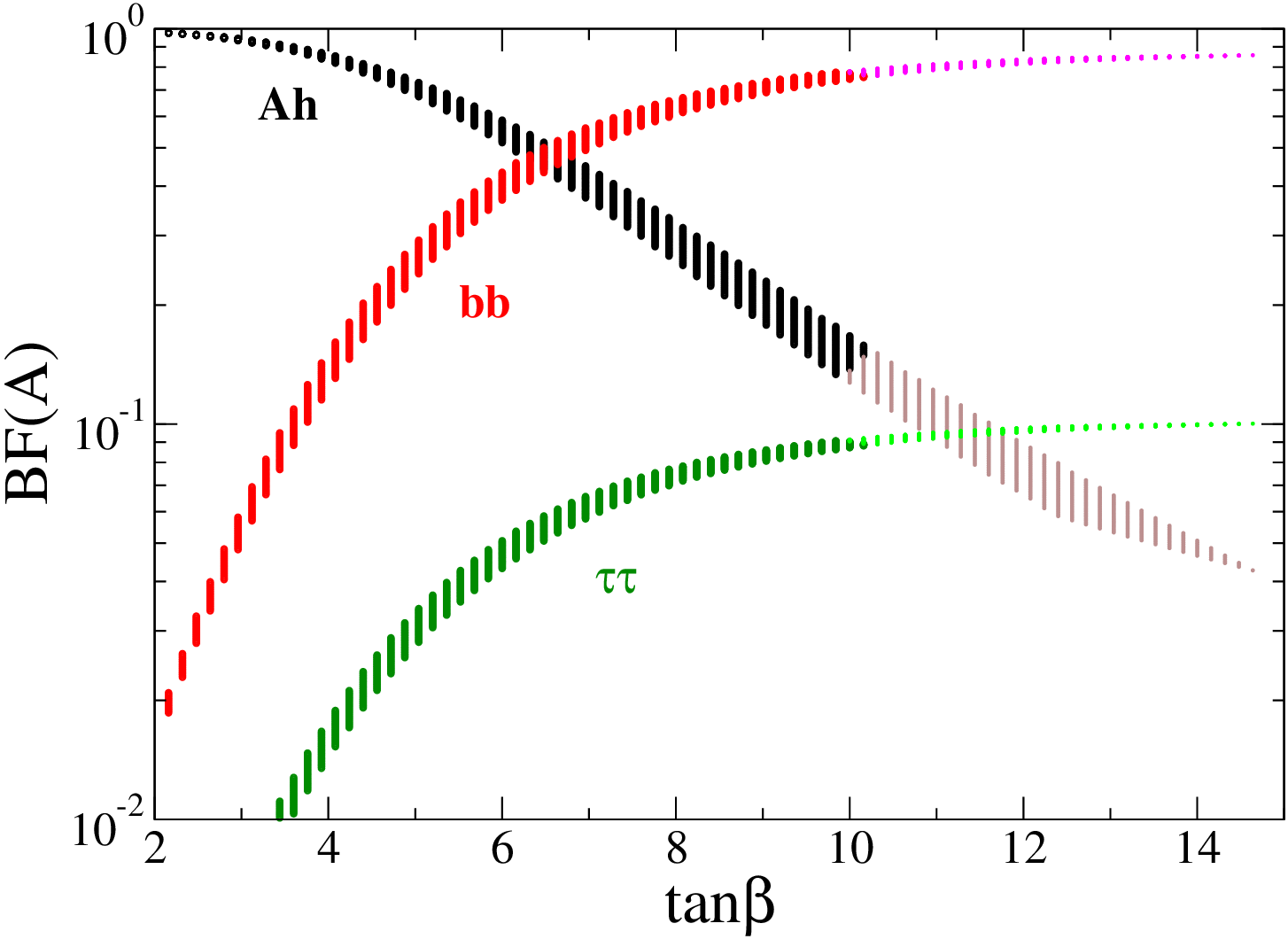}
\caption{Same as Fig.~\ref{decay_H}, but for $A$.}
\label{decay_A}
\end{center}
\end{figure}

\begin{figure}[t]
\begin{center}
\includegraphics[width=75mm]{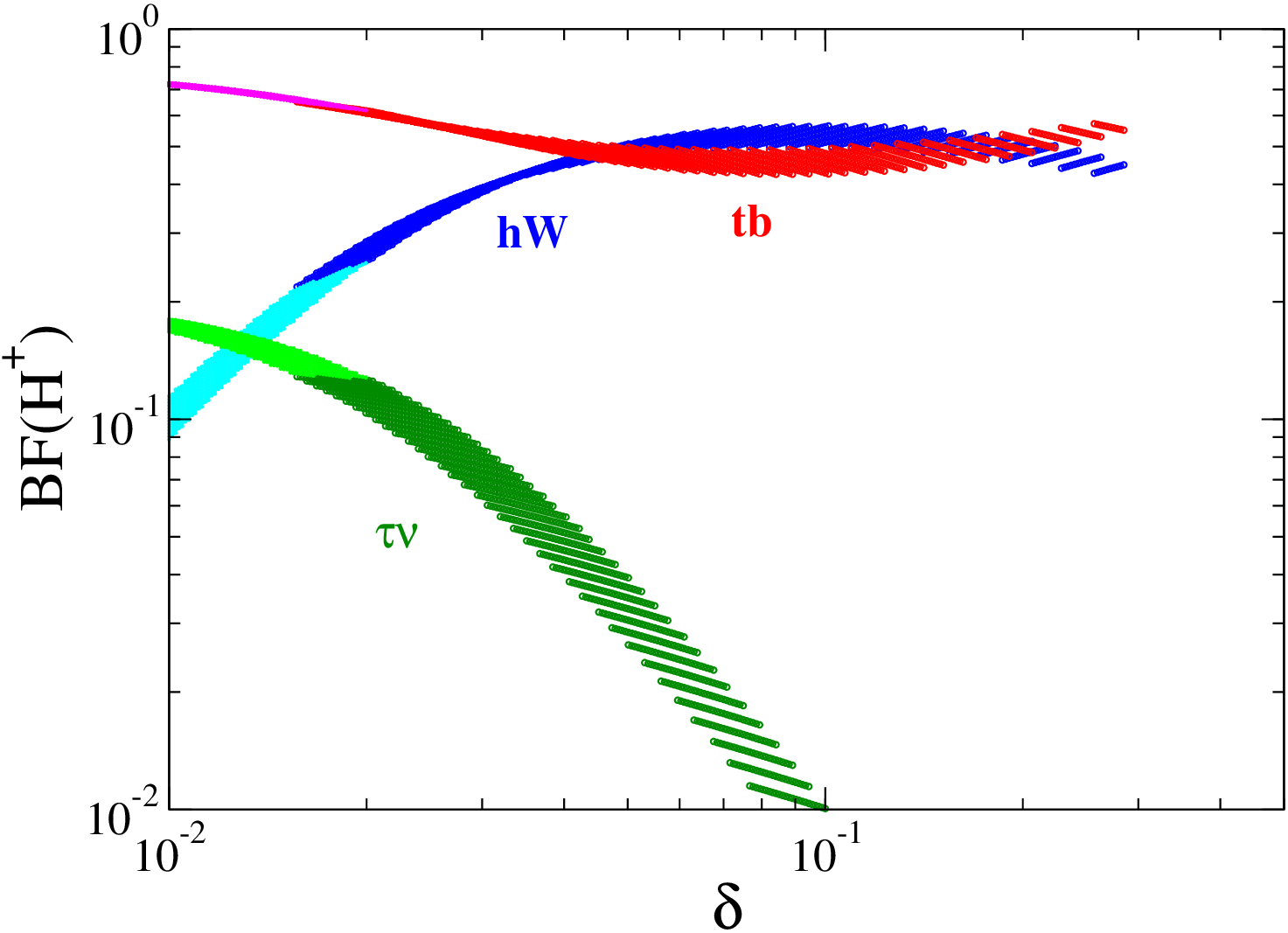}\hspace{5mm}
\includegraphics[width=75mm]{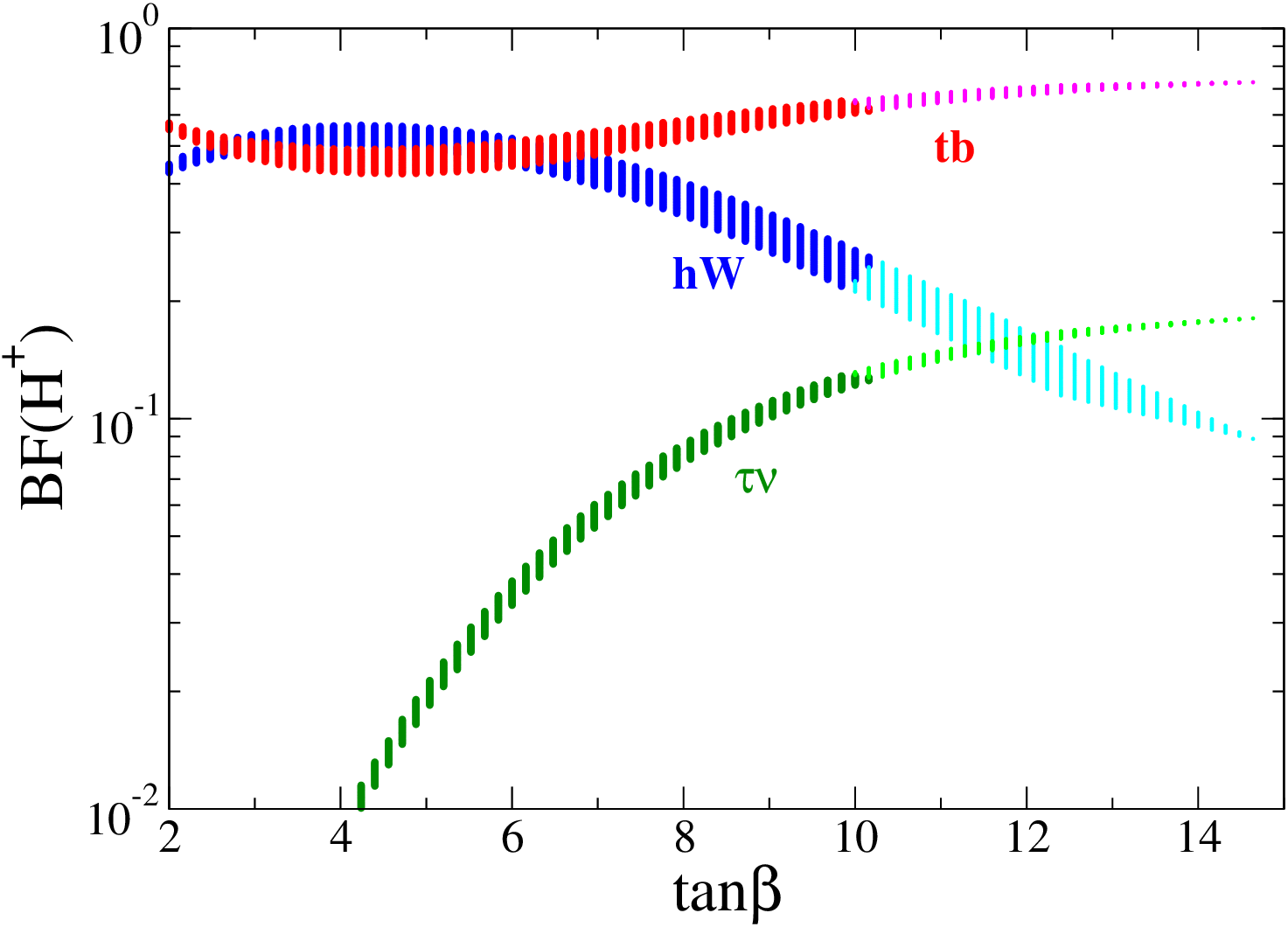}
\caption{Same as Fig.~\ref{decay_H}, but for $H^\pm$.}
\label{decay_ch}
\end{center}
\end{figure}

First, we evaluate the decay branching fractions of $H$, $A$ and $H^\pm$. 
Fig.~\ref{decay_H} shows the decay branching fractions of $H$ as a function of $\delta$ (left panel) and $\tan\beta$ (right panel) in the Type-II THDM.  
The band for each curve indicates the uncertainty in the parameter choice at the 68\% CL.  Note that we have restricted ourselves to the region with $\delta \agt 10^{-2}$ and $2 \alt \tan\beta\alt 10$ in the upper-right panel of Fig.~\ref{contours}.
The parts with light colors are excluded by the heavy neutral Higgs boson search at the LHC~\cite{MSSM_neutral_CMS}. 
It is seen that the gauge boson pair decays of $H$ ($H\to WW$ and $H\to ZZ$) 
become more important than the fermionic decays ($H\to b\bar{b}$ and $H\to \tau\tau$) 
when the value of $\delta$ ($\tan\beta$) is taken to be larger (smaller). 
The situation where $H$ mainly decays into the gauge boson pairs does not happen in the MSSM, for $\delta$ is suppressed in the large $m_A$ regime. 
For example, when $m_A$ is taken to be 300 GeV, the value of $\delta$ is smaller than $10^{-2}$~\cite{Djouadi2}, so that only the fermionic decays of $H$ dominate. 

Fig.~\ref{decay_A} shows the decay branching fractions of $A$ as a function of $\delta$ (left panel) and $\tan\beta$ (right panel) in the same setup as in Fig.~\ref{decay_H}. 
The decay rate of $A\to hZ$ ($A\to b\bar{b}$) is proportional to $\cos(\beta-\alpha)^2\simeq 2\delta$~ ($\tan^2\beta$), 
so that the magnitude of the branching fraction of $A\to hZ$ ($A\to b\bar{b}$) increases (decreases) when $\delta$ gets larger values. 
When $\delta\gtrsim 0.04$ or $\tan\beta \lesssim 7$, the branching fraction of $A\to hZ$ is larger than that of the $A\to b\bar{b}$ decay.

In Fig.~\ref{decay_ch}, the decay branching fractions of $H^\pm$ are shown as a function of $\delta$ (left panel) and $\tan\beta$ (right panel) in the same setup as in Fig.~\ref{decay_H}. 
In the region of $\delta>10^{-2}$, the $H^+\to t\bar{b}$ and $H^\pm\to hW^\pm$ modes are dominant. 

Next, we discuss the production of the extra Higgs bosons.  The extra neutral Higgs bosons $H$ and $A$ are mainly produced via the gluon fusion process: $gg\to H/A$. 
The production cross section is given by
\begin{align}
\sigma(gg\to H/A)=\sigma(gg\to h_{\text{SM}})\times 
\frac{m_{h_{\text{SM}}}^3}{m_{H/A}^3}\frac{\Gamma(H/A\to gg)}{\Gamma(h_{\text{SM}}\to gg)},
\end{align}
where $\sigma(gg\to h_{\text{SM}})$ is the gluon fusion production cross section of the SM Higgs boson $h_{\text{SM}}$ and $\Gamma(h_{\text{SM}}\to gg)$ is the decay rate of $h_{\text{SM}}\to gg$.  We note that the vector boson fusion production mechanism is not useful to produce $H$ and $A$ because the cross section for $H$ is proportional to $\cos(\beta-\alpha)^2\simeq 2\delta$ and that for $A$ is zero at the tree level due to the absence of the $VVA$ vertex.

\begin{figure}[t]
\begin{center}
\includegraphics[width=75mm]{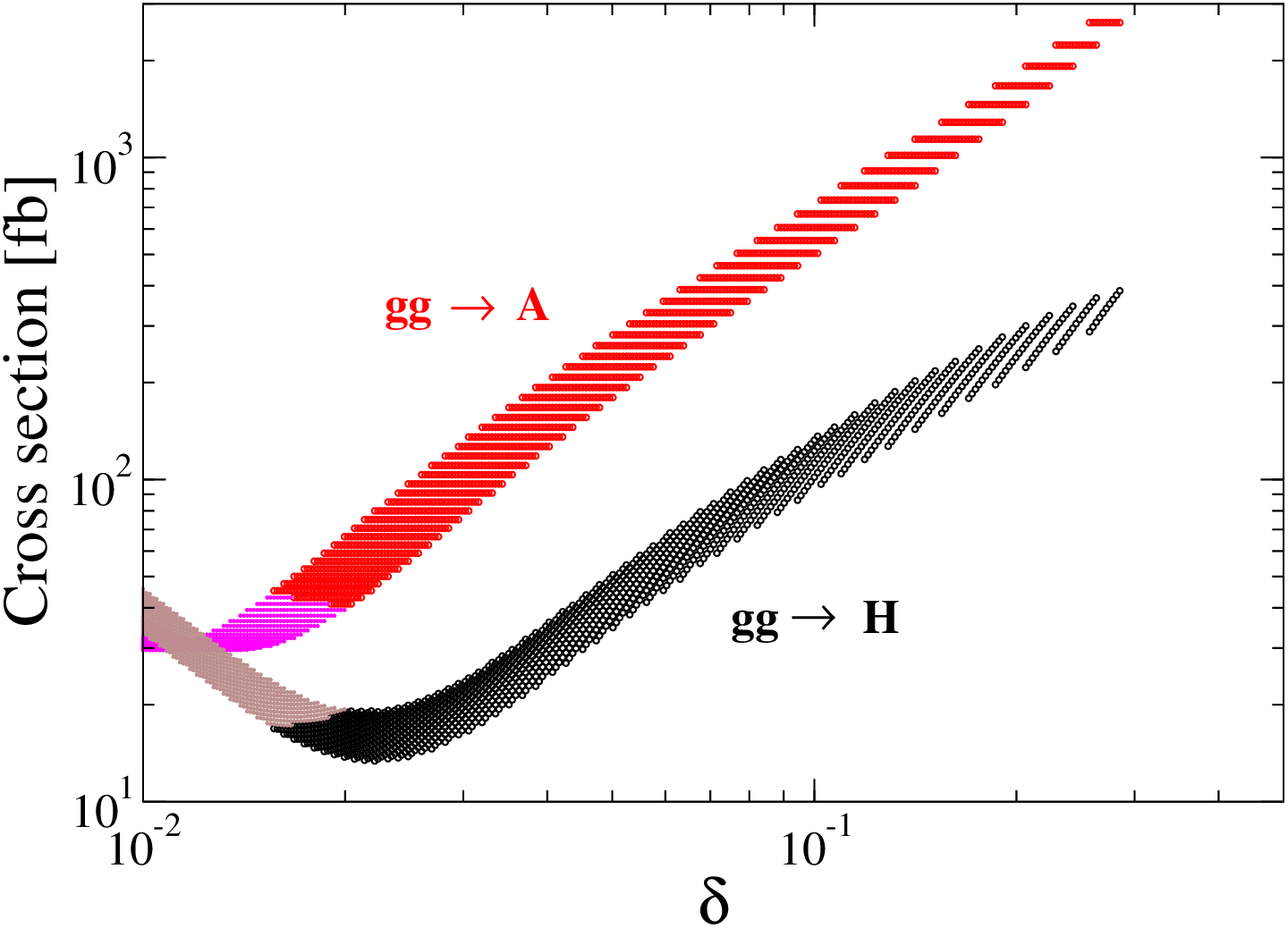} \hspace{5mm}
\includegraphics[width=75mm]{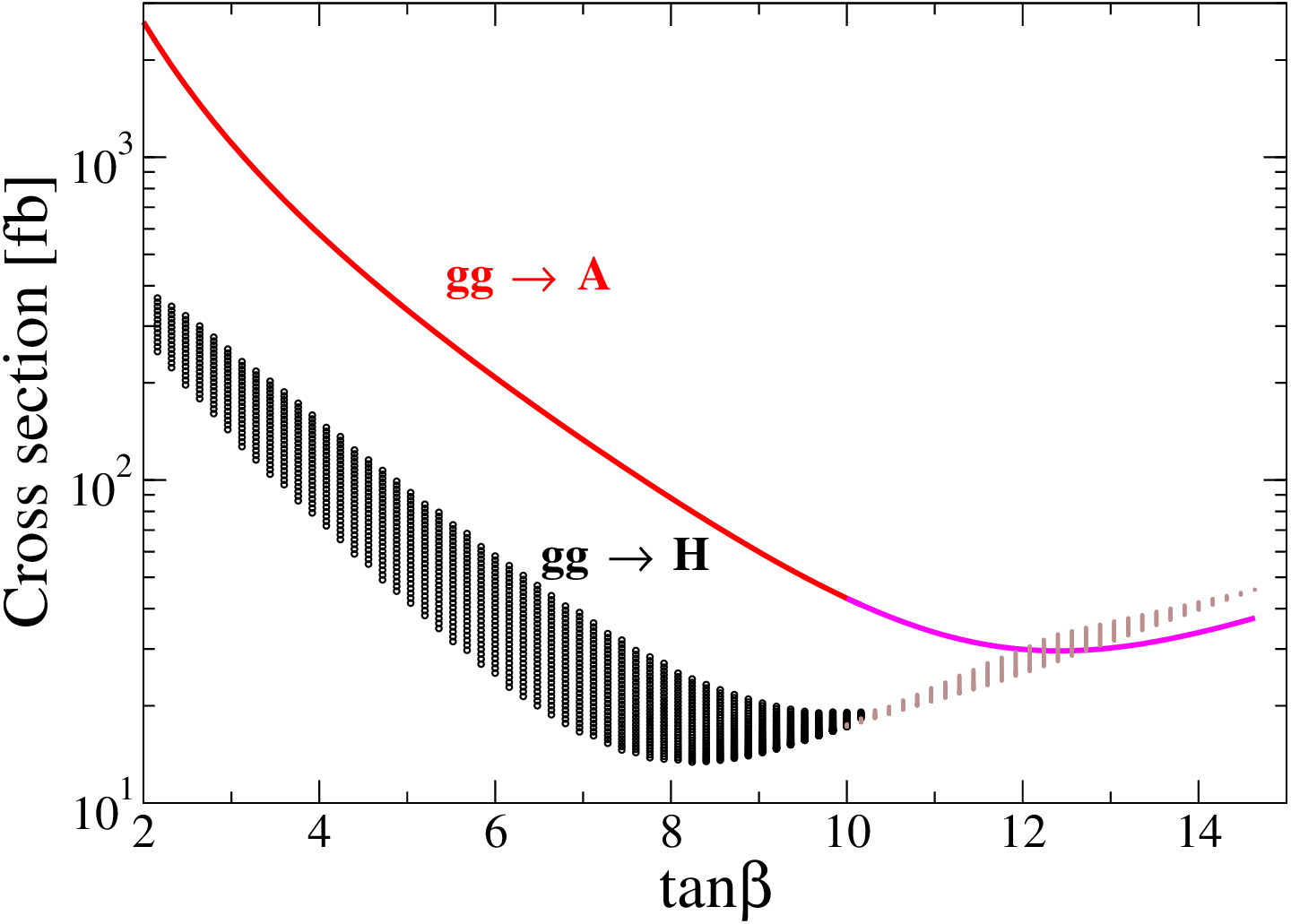}
\caption{Gluon fusion production cross sections for $H$ and $A$ in units of fb as a function of $\delta$ (left panel) and $\tan\beta$ (right panel) for $m_\Phi = M=300$ GeV in the Type-II THDM. 
The collision energy is assumed to be 8 TeV.  The band for each curve reflects parameter uncertainty at the 68\% CL as determined by the Higgs data. 
The light-colored parts are excluded by the LHC search data~\cite{MSSM_neutral_CMS}.}
\label{gfusion_HA}
\end{center}
\end{figure}

In Fig.~\ref{gfusion_HA}, the gluon fusion production cross sections for $H$ and $A$ are shown as a function of $\delta$ (left panel) and $\tan\beta$ (right panel) for the collision energy to be 8 TeV in the Type-II THDM. 
The cross sections are calculated using the gluon fusion cross section of a fictitious 300-GeV Higgs boson in the SM, whose value is 3.606~pb~\cite{gfusion}.
The band for each curve reflects parameter uncertainty at the 68\% CL as determined by the Higgs data. 
The predictions shown by light colors are excluded by the LHC search data~\cite{MSSM_neutral_CMS}. 
Except for $\delta\simeq 10^{-2}$, the cross section of $A$ is larger than that of $H$ because the $Ht\bar{t}$ coupling is proportional to $\sim(\sqrt{2\delta}-\cot\beta)$, so that larger values of $\delta$ give smaller coupling constants, 
while the $At\bar{t}$ coupling is simply proportional to $\cot\beta$. 
No dependence of $\delta$ in the $Af\bar{f}$ couplings can also be seen in the cross section of $A$ in the right plot of Fig.~\ref{gfusion_HA}; namely, the $gg \to A$ cross section is given by a curve without a band.

\begin{figure}[t]
\begin{center}
\includegraphics[width=75mm]{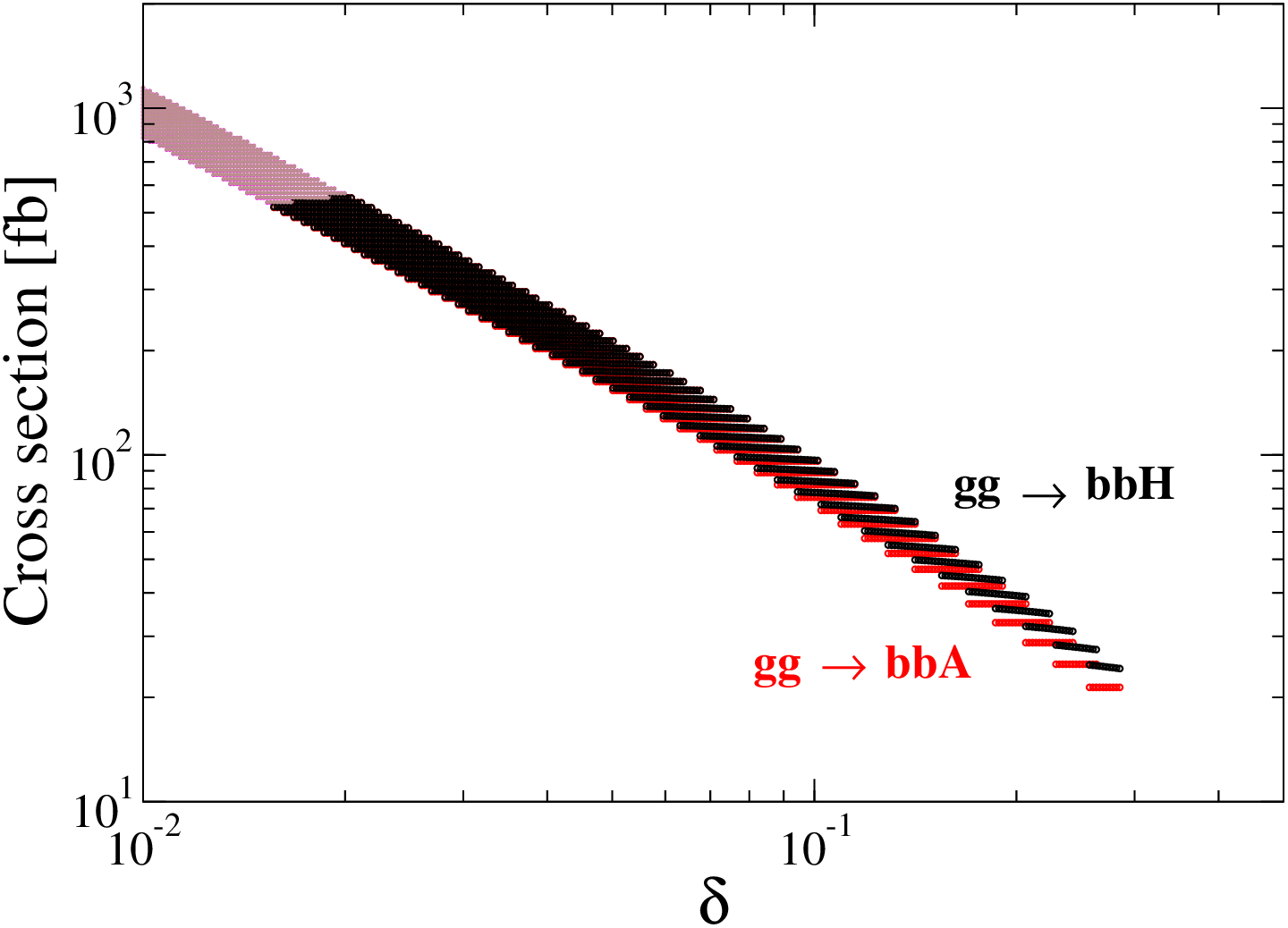} \hspace{5mm}
\includegraphics[width=75mm]{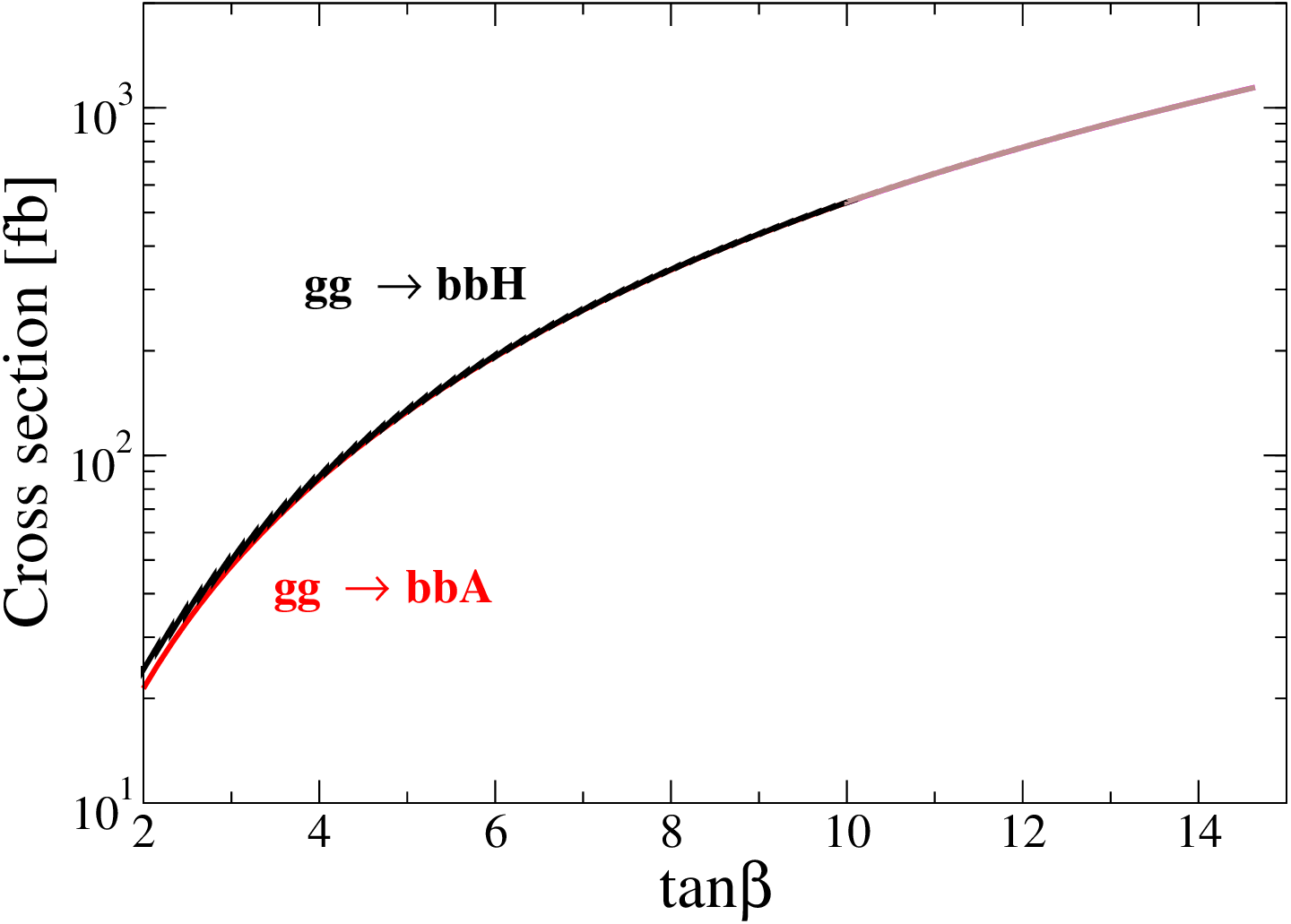}
\caption{Same as Fig.~\ref{gfusion_HA}, but for the bottom quark associate production cross section. }
\label{associate_HA}
\end{center}
\end{figure}

As another single production mechanism for $H$ and $A$, the bottom quark associate processes: $gg\to b\bar{b} H/A$ can be important, especially in the case of large $\tan\beta$. 
The cross sections of these processes are proportional to $(\sqrt{2\delta}+\tan\beta)^2$ and $\tan^2\beta$ for $H$ and $A$, respectively, in the Type-II (and Type-Y) THDM. 

In Fig.~\ref{associate_HA}, the bottom quark associate production cross sections for $H$ and $A$ are plotted as a function of $\delta$ (left panel) and $\tan\beta$ (right panel) in the same setup as in Fig.~\ref{gfusion_HA}. 
The cross sections are calculated by scaling from that with $\tan\beta=1$ and $m_\Phi=300$ GeV whose value is obtained as 5.34~fb using the {\tt MadGraph} package~\cite{MG5} and the {\tt CTEQ6L} parton distribution functions (PDF's).  The cross section for $H$ is almost the same as that for $A$ as they have almost the same scaling behavior, and the maximum value is about 500~fb when $\tan\beta$ is taken to be 10. 

\begin{figure}[t]
\begin{center}
\includegraphics[width=75mm]{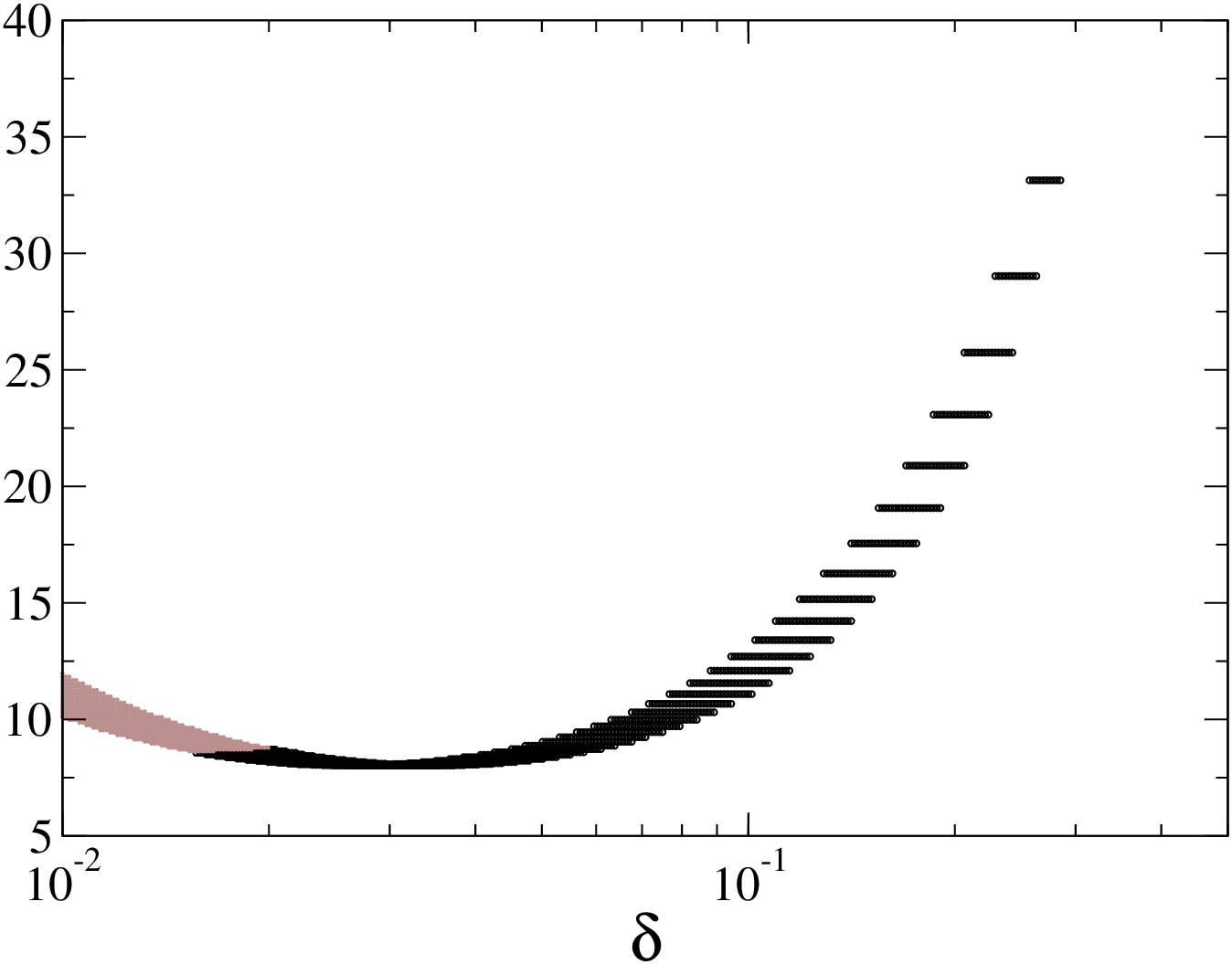} \hspace{5mm}
\includegraphics[width=75mm]{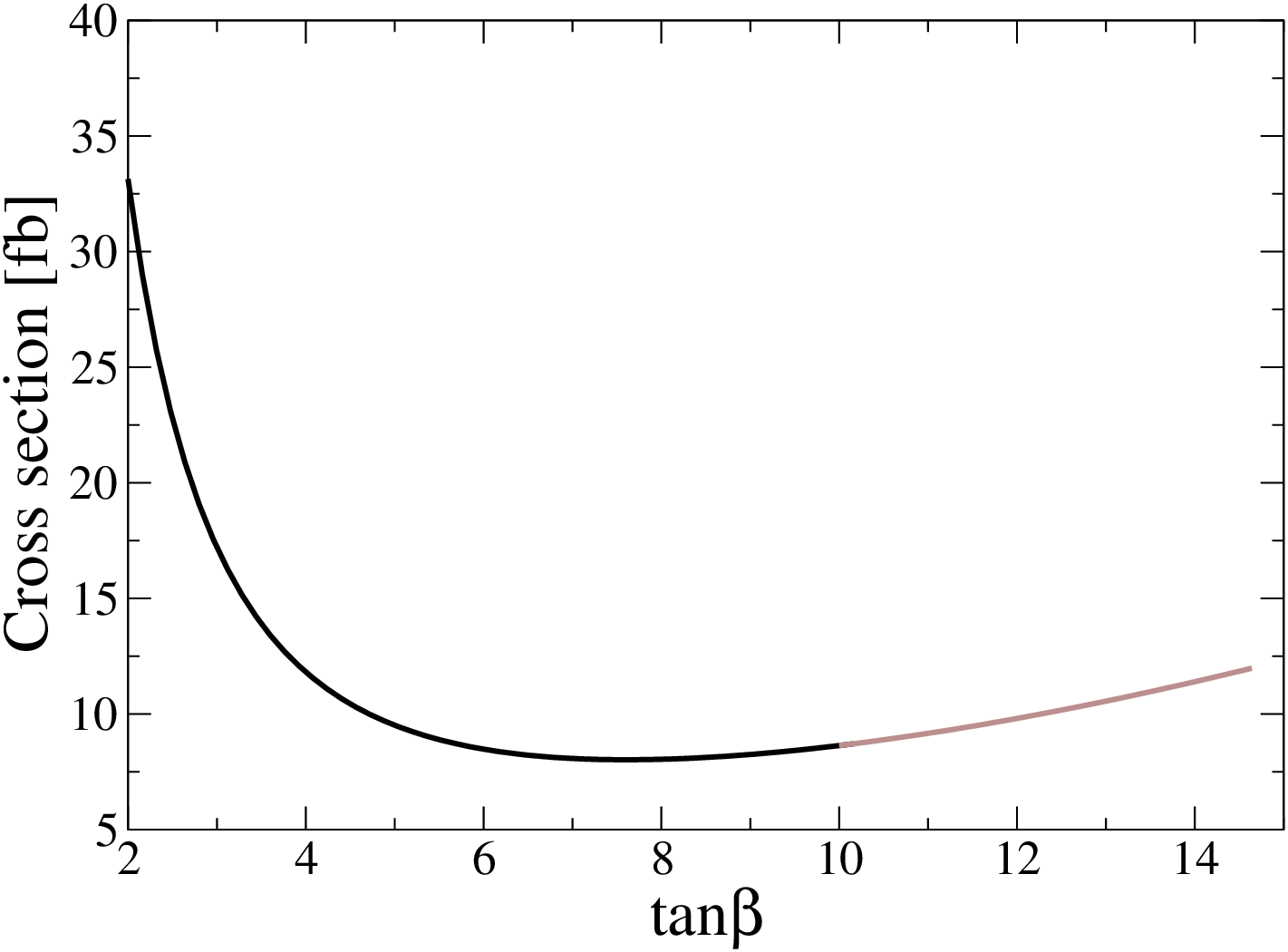}
\caption{Same as Fig.~\ref{gfusion_HA}, but for the $gb\to H^\pm t$ process. }
\label{gb}
\end{center}
\end{figure}

Regarding the charged Higgs boson production, the $gb\to H^\pm t$ process is important, 
especially when the charged Higgs boson mass is larger than the top quark mass. 
The $tbH^\pm$ vertex is proportional to $m_b\tan\beta+m_t\cot\beta$, so that the cross section reaches the minimum when $\tan\beta=\sqrt{m_t/m_b}\simeq 7.5$. 

In Fig.~\ref{gb}, the $gb\to H^\pm t$ cross section is shown as a function of $\delta$ (left panel) and $\tan\beta$ (right panel) in the same setup as in Fig.~\ref{gfusion_HA}. 
The cross section is calculated by scaling from that with $\tan\beta=1$ and $m_\Phi=300$ GeV whose value is obtained as 120~fb using the {\tt MadGraph} package and the {\tt CTEQ6L} PDF's. 
The cross section has a minimum of $\sim 7$ fb when $\tan\beta \simeq 7.5$. 
The maximum value of the cross section is about 33 fb when the $\tan\beta \simeq 2$, corresponding to $\delta\sim 0.3$.

There are also pair production processes for the extra Higgs bosons 
via the $s$-channel gauge boson mediation such as 
$q\bar{q}\to \gamma^*/Z^*\to H^+H^-$, $q\bar{q}\to Z^*\to HA$, $q\bar{q}'\to W^{\pm *}\to HH^\pm$ and $q\bar{q}'\to W^{\pm *}\to AH^\pm$. 
However, the cross sections are suppressed with increasing masses of the extra Higgs bosons. 
For example, when $m_{\Phi}=300$ GeV, the cross section of $pp\to AH^+$ is about 1.5 fb at the collision energy of 8 TeV, and the other cross sections are even smaller~\cite{Djouadi2}.

Finally, we would like to discuss the signal events for the extra Higgs bosons at the LHC in the favored parameter region.  As seen in Figs.~\ref{decay_H}, \ref{decay_A} and \ref{decay_ch}, 
$H$ ($A$) mainly decays into the gauge boson pairs $W^+W^-$ and $ZZ$ ($h$ and the $Z$ boson), while $H^\pm$ mainly decay into $hW^\pm$ and $tb$ in the large $\delta$ region ($\delta \gtrsim 10^{-2}$).  As an example, we here consider the case with $\delta=0.2$ in which the 
allowed range of $\tan\beta$ at the 68\% CL is $2.3<\tan\beta<2.8$. 

\begin{table}[t]
\begin{center}
\begin{tabular}{cccccc}\hline\hline
BF& $\mathcal{B}(H\to W^+W^-)$ & $\mathcal{B}(H\to ZZ)$ & $\mathcal{B}(A\to hZ)$ & $\mathcal{B}(H^\pm\to hW^\pm)$&$\mathcal{B}(H^\pm\to tb)$ \\\hline
[\%]    & 65-68 & 29-30& 95-97 & 45-53 & 47-55 \\\hline
CS & $\sigma(gg\to H)$ & $\sigma(gg\to A)$ & $\sigma(gg\to b\bar{b}H)$ & $\sigma(gg\to b\bar{b}A)$ & $\sigma(gb\to H^-t)$ \\\hline
[fb]&211-300 & 1285-1921& 32-43 & 29-42&19-26 \\\hline\hline
\end{tabular} 
\end{center}
\caption{Branching fractions (BF) and cross sections (CS) of $H$, $A$ and $H^\pm$ for $\delta =0.2$ and $m_\Phi=M=$300 GeV.  The allowed range of $\tan\beta$ at the 68\% CL, $2.3<\tan\beta<2.8$, is reflected in the range of each quantity.} 
\label{bench_mark}
\end{table}

TABLE~\ref{bench_mark} lists the branching fractions and production cross sections of $H$, $A$ and $H^\pm$.  For $H$ ($A$), the gluon fusion production cross section is about 7 (50) times larger than that of the corresponding bottom quark associate production process. 
The $gb\to H^\pm t$ production cross section is about $19-26$ fb. 
The following final states of the signal processes are important to test the scenario with large values of $\delta$:
\begin{align}
gg&\to H\to W^+W^-/ZZ\to \ell^+\ell^-E_T\hspace{-4.5mm}/\hspace{2mm}/\ell^+\ell^-\ell^+\ell^-~~~\text{for }H,\\
gg&\to A\to hZ\to b\bar{b}\ell^+\ell^-~~~\text{for }A,\\
gb&\to H^- t\to hW^-bW^+\to \ell^+\ell^- bb\bar{b} E_T\hspace{-4.5mm}/\hspace{2mm} ~~~\text{for } H^\pm, 
\end{align}
where $\ell^\pm$ represent $e^\pm$ or $\mu^\pm$. 
The signal processes $gg\to H\to W^+W^-/ZZ$ have the main backgrounds from $pp\to W^+W^-/ZZ$, and are the same as the SM Higgs boson with a fictitious mass of 300 GeV. 
For the signal event $gg\to A\to hZ$ ($gb\to tH^-$), 
the $pp\to ZZ$ and $q\bar{q}\to Z^* \to h_{\text{SM}}Z$ ($gb\to W^+W^-h_{\text{SM}}b$) processes can be main backgrounds. 

In TABLE~\ref{sb}, the signal and background cross sections are listed assuming the LHC collision energy to be 8 TeV.  We here assume that the b-jets can be identified with 100\% efficiency.  Except for the signal events from $H$ (first two columns), the other two cross sections for the signal events are comparable or larger than those of the background. 
The cross sections for the signal events from $H$ is smaller by about two order of the magnitude than that from the backgrounds. 
However, for the final states of $\ell^+\ell^-E_T\hspace{-4.5mm}/\hspace{2mm}$ and $\ell^+\ell^-\ell^+\ell^-$ coming from the $H$ signal processes, the transverse mass distribution~\cite{transverse_mass} and the invariant mass distribution of the $\ell^+\ell^-\ell^+\ell^-$ system can be useful in increasing the signal-to-background ratio. 
The Jacobian peak (resonance peak) of the signal in the transverse mass distribution (the invariant mass distribution) can be used to determine the mass of $H$, while such a characteristic feature cannot be observed in the background events. 

\begin{table}[t]
\begin{center}
\begin{tabular}{ccccc}\hline\hline
Events~ & ~~$\ell^+\ell^-E_T\hspace{-4.5mm}/\hspace{2mm}$~~ & 
~~$\ell^+\ell^-\ell^+\ell^-$~~ & ~~$b\bar{b}\ell^+\ell^-$~~ & 
~~$\ell^+\ell^-bb\bar{b} E_T\hspace{-4.5mm}/\hspace{2mm}$~ \\\hline
Signal [fb]   & $7.1-9.5$ & $0.29-0.40$ & $24-37$ & $0.29-0.34$  \\\hline
Backg. [fb] & $1.73\times 10^{3}$ & $22.5$ & 67.3 & 0.048 \\\hline\hline
\end{tabular} 
\end{center}
\caption{Cross sections of the signal and background processes for the collision energy of 8 TeV  at the LHC, obtained using the {\tt MadGraph} package and the {\tt CTEQ6L} PDF's. } 
\label{sb}
\end{table}

We now briefly comment on the Higgs phenomenology in the other parameter region preferred by the LHC data, {\it i.e.}, the region with $\delta\lesssim 10^{-2}$ in Fig.~\ref{contours}. 
In that case, the $HVV$, $AhZ$ and $H^\pm hW^\mp$ couplings are much suppressed, and the decay branching fractions of $H\to VV$, $A\to hZ$ and $H^\pm \to hW^\mp$ are negligible.  Instead of such gauge boson associate decay modes, these heavy Higgs bosons mainly decay into fermion pairs whose pattern depends on the type of Yukawa interactions, the masses of the extra Higgs bosons, and $\tan\beta$. 
For example, $H$ and $A$ can decay dominantly into $b\bar{b}$ ($\tau\tau$) for large $\tan\beta$ in the Type-I, Type-II and Type-Y (Type-X) THDM when the masses of both $H$ and $A$ are smaller than $2m_t$. 
On the other hand, the main decay mode of $H^\pm$ can be $\tau\nu$ ($cs$ and $cb$) for large $\tan\beta$ in the Type-I, Type-II and Type-X (Type-Y) THDM when the charged Higgs boson mass is smaller than the top quark mass~\cite{typeX}.


\section{Conclusions \label{sec:summary}}

In this work, we have considered the two-Higgs doublet models (THDM's) with softly broken $Z_2$ symmetry introduced to avoid flavor-changing neutral currents at the tree level and the four independent scenarios (Type-I, Type-II, Type-X and Type-Y) differing in the Yukawa interactions. 
We have scanned the parameter regions that can explain the current Higgs boson search data recently reported by the ATLAS and CMS Collaborations.  We have found that the Type-II THDM can best explain the data among all, with two separate parameter regions on the $\tan\beta$-$\delta$ plane at the 68\% CL.  One region has smaller $\delta$ ($< 10^{-2}$), and the other implies the relations $c_{hbb}$, $c_{h\tau\tau}\simeq -1$.  The latter case, with $\tan\beta \simeq \sqrt{2/\delta}$ and $\delta > 10^{-2}$, is ignored in previous analyses.  Based on this finding, we have then studied the phenomenology of the extra Higgs bosons in this region.  The heavy CP-even Higgs boson primarily decays into a pair of $W$ and $Z$ bosons, different from the corresponding particle in the minimal supersymmetric standard model.  The CP-odd Higgs boson mainly decays to the SM-like Higgs and $Z$ bosons, and the charged Higgs boson to the SM-like Higgs and $W$ bosons or a pair of top and bottom quarks.  
We have also computed the cross sections of various signal and background events for the production of the extra Higgs bosons.

\noindent
\section{acknowledgments}

This research was supported in part by the National Science Council of Taiwan, R.~O.~C.
under Grant Nos.~NSC-100-2628-M-008-003-MY4 and NSC-101-2811-M-008-014.

\bigskip
\noindent
{\it Note added:}

While this manuscript was being written up, we noticed the appearance of
Ref.~\cite{Celis:2013rcs} in which a global fit to the Higgs boson search data at the LHC was discussed for various THDM's, including the softly broken $Z_2$ symmetric THDM's considered in this work.

\end{document}